\documentclass[a4paper,11pt]{article}

\usepackage{amssymb}
\usepackage{amsmath}
\usepackage{graphicx}
\usepackage{color}
\usepackage[a4paper]{geometry}
\usepackage[numbers]{natbib}
\usepackage{caption}
\usepackage{setspace}
\usepackage{epstopdf}
\usepackage{hyperref}
\usepackage{float}
\usepackage{booktabs}
\usepackage{multirow}
\usepackage{enumerate}
\usepackage{bm}
\usepackage{subfig}

\hypersetup{
	pdftitle={Geometrical nonlinear 2D CUF models},    
	pdfauthor={mul2},    
	pdfsubject={Geometrically nonlinear analysis of laminated composite plates based on refined finite element models},   
	pdfkeywords={Higher-order plate elements; Unified Formulation; Geometrical nonlinearities}, 
	colorlinks=true,       
	linkcolor=blue,          
	citecolor=red,        
	filecolor=magenta,      
	urlcolor=blue           
}
\newcommand{\vect}[1]{\boldsymbol{\mathbf{#1}}}

\title{Buckling and post-buckling of anisotropic flat panels subjected to axial and shear in-plane loadings  accounting for classical and refined structural and nonlinear theories}
\author{
E. Carrera$^{1,2}$\footnote{Professor of Aerospace Structures and
Aeroelasticity. E-mail: erasmo.carrera@polito.it},
R. Azzara$^{1}$\footnote{PhD student. E-mail: rodolfo.azzara@polito.it},
E. Daneshkhah$^{1}$\footnote{PhD student. E-mail: ehsan.daneshkhah@polito.it},
A. Pagani$^{1}$\footnote{Associate professor. E-mail: alfonso.pagani@polito.it},
B. Wu$^{3}$\footnote{Marie Curie Individual Fellow. E-mail: wubinlongchang@163.com}
\\ \\
$^{1}${\em Mul}$\,^2$ group \\
Department of Mechanical and Aerospace Engineering, Politecnico di Torino, \\
Corso Duca degli Abruzzi 24, 10129 Torino, Italy\\\\
$^2$Department of Mechanical Engineering, College of Engineering, \\ Prince Mohammad Bin Fahd University P.O. Box 1664. \\ Al Khobar 31952. Kingdom of Saudi Arabia. \\\\ $^{3}$School of Mathematics, Statistics and Applied Mathematics, \\ NUI Galway, University Road, Galway, Ireland 
}

\date{}

\begin{document}

\maketitle
	
\noindent
{\bf Abstract:} 
{\em
This article investigates the large deflection and post-buckling of composite plates by employing the Carrera Unified Formulation (CUF). As a consequence, the geometrically nonlinear governing equations and the relevant incremental equations are derived in terms of fundamental nuclei, which are invariant of the theory approximation order. By using the Lagrange expansion functions across the laminate thickness and the classical finite element (FE) approximation, layerwise (LW) refined plate models are implemented. 
The Newton-Raphson linearization scheme with the path-following method based on the arc-length constraint is employed to solve geometrically nonlinear composite plate problems.
In this study, different composite plates subjected to large deflections/rotations and post-buckling are analyzed, and the corresponding equilibrium curves are compared with the results in the available literature or the traditional FEM-based solutions. The effects of various parameters, such as stacking sequence, number of layers, loading conditions, and edge conditions are demonstrated. The accuracy and reliability of the proposed method for solving the composite plates' geometrically nonlinear problems are verified. 
}
\hrule
\section{Introduction} 
The availability of new constituents and manufacturing processes has widely promoted the use of composite laminated structures, which are now constantly employed in aeronautics, aerospace, automotive and construction engineering \cite{kaw2005mechanics}. Understanding the mechanical behaviours of composites is, therefore more than ever, a major concern for designers and researchers \cite{forouzan2019damage}. 

Although stiff, composite laminates can be designed to carry out large elastic displacements and rotations when sufficiently thin. It is the case, for example, of coilable composite tape springs, see for example \cite{Pagani2020d}. In this context, new research trends are arising for characterizing the geometrically nonlinear mechanics of such laminates \cite{dash2015static,singh2008nonlinear}. 
Zhang and Yang \cite{zhang2009recent} provided a review of recent developments in finite element (FE) analysis for laminated plate structures. 
Han \textit{et al.} \cite{han1994investigation} extended the hierarchical FE method to geometrical nonlinearities for studying the response of composite plates. By using first-order shear deformation theory (FSDT) and adopting the total Lagrangian formulation, Zhang and Kim \cite{zhang2005simple} performed nonlinear analysis of different laminated structures, employing a displacement-based 3-node triangular plate/shell element. The same authors also developed a displacement-based 4-node quadrilateral element for geometrically nonlinear analysis of laminated plates \cite{zhang2006geometrically}. A numerical method based on the higher-order shear deformation theory (HSDT) and isogeometric analysis was provided by Tran \textit{et al.} \cite{tran2015geometrically} to investigate geometrically nonlinear response of two-dimensional (2D) laminated structures. 
Sridhar and Rao \cite{sridhar1995large} utilized the four-node quadrilateral composite shell FE to analyze laminated circular composite plates in the regime of large-displacement fields. Liew \textit{et al.} \cite{liew2004nonlinear} employed a mesh-free kp-Ritz method combined with the FSDT for the nonlinear flexural analysis of plates. A geometrically nonlinear parametric instability of functionally graded plates was studied by  Alijani and Amabili \cite{alijani2013non} adopting a multi-degree-of-freedom Lagrangian formulation and nonlinear higher-order shear deformation theory.
Nonlinear static analysis of composite thick plates resting on nonlinear elastic foundations was conducted by Baltac{\i}o{\u{g}}lu \textit{et al.} \cite{baltaciouglu2011large} based on a discrete singular convolution approach. Reddy \textit{et al.} \cite{reddy2012bending} studied the effect of different geometrical and loading parameters on the bending analysis of laminated composite plates. Readers are referred to  \cite{reddy2003mechanics,gorji1986large,zhang1985large} for more works on the nonlinear response analysis of the large displacements/rotations of composite plates.
A complete description of nonlinear vibrations and stability of shell and plates structures was provided by Amabili \cite{amabili2008nonlinear,amabili2018nonlinear}.

Among the problems characterized by geometrically nonlinear response, the post-buckling phenomenon of composite plates deserves special attention. Many researchers have addressed this topic indeed \cite{hui1990buckling,hui2017postbuckling,zhang1983postbuckling,dash2012buckling}. For example, Leissa \cite{leissa1987review} presented a review of 2D laminated composite plate buckling analysis.
In addition, the book edited by Turvey and Marshal \cite{turvey2012buckling} conducted comprehensive studies on buckling and post-buckling of composite plates. Librescu and Stein \cite{librescu1991geometrically} formulated a geometrically nonlinear theory of isotropic symmetrically laminated plates and analyzed their post-buckling behaviours. The effect of material nonlinearity on the post-buckling of composite plates and shells was studied by Wang and Srinivasan \cite{wang1995effect}. Sundaresan \textit{et al.} \cite{sundaresan1996buckling} described the buckling and post-buckling behaviours of typically 2D thick laminated rectangular plates, where they developed a eight-node isoparametric plate FE. Liew \textit{et al.} \cite{liew2006postbuckling} proposed a Ritz method combined with the FSDT and kernel particle approximation for the field variables to investigate the post-buckling behaviour of 2D laminated structures. 
Amabili and Tajahmadi \cite{amabili2012thermal} performed post-buckling analysis of isotropic and composite plate structures subjected to thermal changes. 
Chen and Qiao \cite{chen2014post} carried out a post-buckling analysis of 2D composite plates subjected to combined compressive and shear loadings using the finite strip method. 
Dash and Singh \cite{dash2014buckling} performed buckling and post-buckling response analyses of laminated plates with random system properties. 
 
In the present paper, the Carrera Unified Formulation (CUF), which has been proved to be an efficient and accurate method for solving nonlinear structural problems \cite{Pagani2018, PaganC_CS_2017, wu2019accurate, wu2019large,petrolo2018global,kaleel2019computationally,kaleel2019effectiveness}, is now extended to deal with geometrically nonlinear analysis of composite plates. 
In this work, the layerwise (LW) approach based on Lagrange expansions is employed (see \cite{Carre_AJ_1999,carrera1999multilayered2} for an exhaustive derivation of CUF in LW framework). This method provides us with interface compatibility conditions to be easily imposed between different layers. Moreover, by using the CUF, the expansions along the thickness could be selected of any arbitrary order.
As demonstrated in \cite{Pagani2018}, the advantage of CUF is that the nonlinear equilibrium and incremental equations are written in terms of \textit{fundamental nuclei} (FNs) of the secant and tangent stiffness matrices. FNs are invariant of the theory approximation order, thus lower- to higher-order and eventually layerwise CUF structural models can be formulated with ease \cite{PaganC_CS_2017}.
In the same framework, Pagani \textit{et al.} \cite{pagani2019evaluation} evaluated the effect of different geometrically nonlinear models on the response of thin-walled structures adopting a refined beam model. Pagani \textit{et al.} \cite{pagani2020evaluation} utilized a refined CUF plate model to investigate the effect of various strain-displacement nonlinear approximations on the large deflection and post-buckling of isotropic plates. Readers can be referred to \cite{carrera2020vibration,wu2019geometrically,pagani2020accurate} for important works on nonlinear analyses of 2D isotropic and composite shell structures in the CUF framework.

The present research aims to use the CUF and arc-length method with path-following constraint to study the geometrically nonlinear composite plate problems. In this regard, different composite plates subjected to large displacements/rotations and post-buckling are investigated. Large-deflection and post-buckling problems are solved by using the Newton-Raphson linearization scheme for different symmetric and antisymmetric composite plates. The corresponding equilibrium curves are also compared with the results in the available literature or the traditional FEM-based solutions, and the effects of different parameters such as the stacking sequence, number of layers, loading conditions, and edge conditions are investigated.

This paper is structured as follows: (i) first, preliminary information about the nonlinear geometrical relations are provided in Section \ref{X}, including the 2D CUF plate model adopted; (ii) next, numerical results are reported in Section \ref{V}, and they involve laminated composite plates with different boundary and loading conditions; (ii) finally, conclusions are presented in Section \ref{Z}.
%
%
\section{Geometrically nonlinear unified finite plate element} \label{X}
\subsection{Preliminary considerations} 
\begin{figure}[htbp]
\centering
\includegraphics[width=0.4\textwidth,angle=0]{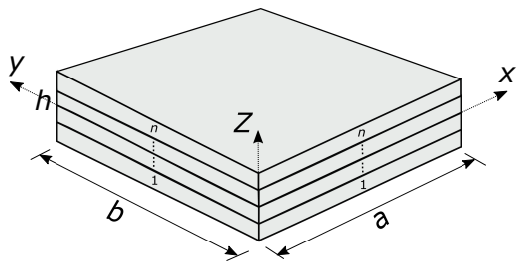}
\hspace{0.00\textwidth}
\caption{The $n$-ply laminated plate and related coordinate reference system.}
\label{comp1}
\end{figure}

In this section, the $n$-ply laminated plate, as illustrated in Fig. \ref{comp1}, is assumed to be located in the $x-y$ plane, whereas the thickness direction lays along $z$-axis. 
The 3D displacement, strain and stress vectors of a given point in the composite plate are defined, respectively as follows:
\begin{equation}
\begin{array}{l}
{\bm{u}}^{k}(x,y,z) = {\{ {u^{k}_{x}}\,{\textrm{ }}{u^{k}_{y}}{\textrm{  }}{u^{k}_{z}}\} ^\text{T}} \\\\  
\bm{\epsilon}^{k} = {\{ {\varepsilon^{k} _{xx}}{\textrm{  }}{\varepsilon^{k} _{yy}}\;{\textrm{ }}{\varepsilon^{k} _{zz}}{\textrm{ }}\;{\varepsilon^{k} _{xz}}\;{\textrm{ }}{\varepsilon^{k} _{yz}}{\textrm{  }}{\varepsilon^{k}_{xy}}\} ^\text{T}} \\\\
{\bm{\sigma }}^{k} = {\{ {\sigma^{k}_{xx}}{\textrm{ }}{\sigma^{k} _{yy}}\,\,{\sigma^{k}_{zz}}\,{\textrm{ }}{\sigma^{k}_{xz}}{\textrm{  }}{\sigma^{k}_{yz}}\,\,{\sigma^{k}_{xy}}\} ^\text{T}}   
\end{array}
\end{equation}
where the superscript $k$ denotes the $k^{th}$-layer of the laminated plate and the superscript $\text{T}$ signifies the transpose.
The presented approach, based on the total Lagrangian formulations, employs the Green-Lagrange strain $\bm{\epsilon}$ and the second Piola-Kirchhoff stress $\bm{\sigma}$, which are work-conjugate. 
The Green-Lagrange strain vector $\bm{{\epsilon}}^{k}$ is obtained via the displacement-strain relations as:
\begin{equation} \label{eq1}
\bm{\epsilon}^{k}= \bm{\epsilon}_{l}^{k}+\bm{\epsilon}_{nl}^{k}= (\bm{b}_{l}+\bm{b}_{nl})\bm{u}^{k}
\end{equation}
in which $\bm{b}_{l}$ and $\bm{b}_{nl}$ represent the 6$\times$3 linear and nonlinear differential operators and they are defined as follows:
\renewcommand{\arraystretch}{0.7}
\begin{equation} \label{eq:bnl}
\bm{b}_l = \left[\begin{array}{ccc}
\partial_x & 0 & 0 \\\\
0 & \partial_y & 0 \\\\
0 & 0 & \partial_z \\\\
\partial_z & 0 & \partial_x \\\\
0 & \partial_z & \partial_y \\\\
\partial_y & \partial_x & 0
\end {array}\right]\text{,} \qquad
\bm{b}_{nl} = \left[\begin{array}{ccc}
 \dfrac{1}{2} \left(\partial_x\right)^2 &  \dfrac{1}{2} \left(\partial_x\right)^2 &  \dfrac{1}{2} \left(\partial_x\right)^2 \\\\
\dfrac{1}{2} \left(\partial_y\right)^2 &   \dfrac{1}{2} \left(\partial_y\right)^2 &  \dfrac{1}{2} \left(\partial_y\right)^2 \\\\
\dfrac{1}{2} \left(\partial_z\right)^2 &   \dfrac{1}{2} \left(\partial_z\right)^2 &  \dfrac{1}{2} \left(\partial_z\right)^2 \\\\
\partial_x \, \partial_z &  \partial_x \, \partial_z &  \partial_x \, \partial_z \\\\
\partial_y \, \partial_z &  \partial_y \, \partial_z &  \partial_y \, \partial_z \\\\
\partial_x \, \partial_y &  \partial_x \, \partial_y &  \partial_x \, \partial_y
\end {array}\right],
\end{equation}
\renewcommand{\arraystretch}{1.}
where $\partial_x = \partial (\cdot)/{\partial x}$, $\partial_y = {\partial (\cdot)}/{\partial y}$, $\partial_z = {\partial (\cdot)}/{\partial z}$ are the partial derivatives.

The following constitutive equation is considered for the stress-strain relationship:
\begin{equation} \label{eq:hooke}
\bm{\sigma}^{k}= \widetilde {\bf{C}}^{k}\bm{\epsilon}^{k}
\end{equation}
where $\widetilde {\bf{C}}^{k}$ is the material elastic matrix. Readers are referred to \cite{Bathe_1996,ochoa1992finite} for the complete expressions of this matrix.
%
%
%
\subsection{Carrera Unified Formulation (CUF)} \label{S}
\subsubsection{Theory kinematics and finite element approximation}
In the framework of the 2D Carrera Unified Formulation \cite{CarreCPZ_2014}, the 3D displacement field ${\bm{u}}(x,y,z)$ is written as a general expansion of the primary unknowns: 
\begin{equation}\label{ftt}
\bm{u}(x,y,z) = F_{s}(z)\bm{u}_{s}(x,y), \qquad s = 0, 1,\ldots,N,
\end{equation}
in which $F_{s}$ represents a set of thickness expansion functions, $\bm{u}_{s}$ indicates the generalized displacement vector depending on the in-plane coordinates $x$ and $y$, $N$ is the order of expansion in the thickness direction and the repeated index $s$ denotes summation \cite{pagani2020evaluation}. In the present work, Lagrange polynomials (LE) are assumed for the expansion functions $F_{s}$ due to the promising and accurate results of the LE \cite{CarrePP_AJ_2013,CarreP_AJ_2016,CarreP_JoSaV_2014}.
It should be noted that the unknown variables are pure displacements in the case of LE. In addition, the LW approach based on the LE is used in this research. This method provides us with interface compatibility conditions to be easily imposed between different layers. By using the CUF, the expansions along the thickness could be selected of any arbitrary order. Therefore, different refined plate models can be obtained by changing the expansion order.
The acronym LDN (Layerwise Displacement-based theory of expansion order N) will be adopted in this article to refer to specific refined plate models. For instance, LD1, LD2, and LD3 indicate linear (two-node), quadratic (three-node), and cubic (four-node) Lagrange expansion functions, respectively, which are utilized to formulate CUF plate models with linear to higher-order kinematics. 

According to the finite element method (FEM), the generalized displacement vector $\bm{u}_{s}$ is approximated based on the FE nodal parameters $\bm{q}_{{s}{j}}$ and shape functions $N_{j}$ as: 
\begin{equation}\label{ut}
\bm{u}_{s}(x,y) = N_{j}(x,y) \bm{q}_{sj}, \qquad j=1,2,\ldots,n_{el},
\end{equation}
in which $N_{j}$ are the $j^{th}$ shape function, $\bm{q}_{sj}$ represents the unknown nodal variables, $n_{el}$ is the number of nodes per element and $j$ indicates summation.
In this article, the classical 2D nine-node quadratic (Q9) FEs are employed for the shape functions in the $x-y$ plane. More details about the Lagrange polynomials and shape functions can be found in \cite{Bathe_1996}.
\subsubsection{Fundamental nuclei of the secant and tangent stiffness matrices} 
The nonlinear governing equations can be derived by using the principle of virtual work, that is,
\begin{equation}\label{eq:equilibrium1}
\delta L_\mathrm{int} = \delta L_\mathrm{ext}
\end{equation}
The virtual variation of internal strain energy in Eq. (\ref{eq:equilibrium1}) is calculated as:
\begin{equation}\label{eq:Lint}
\delta L_\mathrm{int} =
< \delta \vect{\epsilon}^{\textrm{T}} \vect{\sigma} >
\end{equation}
in which $<(\cdot)>=\int_V(\cdot)\;dV$, where $V$ represents the initial undeformed volume of the composite plate structure considering the hypothesis of small deformations, and $\delta$ is the virtual variation operator. According to FEM approximation (\ref{ut}), CUF (\ref{ftt}), constitutive equations (\ref{eq:hooke}) and geometric relations (\ref{eq1}), it is proved that:  
\begin{equation}\label{eq:Lint_2}
\renewcommand{\arraystretch}{2.0}
\begin{array}{lll}
\delta L_\mathrm{int} & = & \delta \vect{q}_{\tau i}^\textrm{T} < \left( \vect{B}_l^{\tau i} + 2 \, \vect{B}_{nl}^{\tau i} \right)^\textrm{T} \vect{C} \left( \vect{B}_l^{sj} + \vect{B}_{nl}^{sj} \right) > \; \vect{q}_{sj} \\
& = &  \delta \vect{q}_{\tau i}^\textrm{T} \, \vect{K}_0^{i j \tau s} \, \vect{q}_{sj} + \delta \vect{q}_{\tau i}^\textrm{T} \, \vect{K}_{lnl}^{i j \tau s} \, \vect{q}_{sj} + \delta \vect{q}_{\tau i}^\text{T} \, \vect{K}_{nll}^{i j \tau s} \, \vect{q}_{sj} + \delta \vect{q}_{\tau i}^\textrm{T} \, \vect{K}_{nlnl}^{i j \tau s} \, \vect{q}_{sj} \\
& = &  \delta \vect{q}_{\tau i}^\text{T} \, \vect{K}_S^{i j \tau s} \, \vect{q}_{sj}
\end{array}
\end{equation}
where the two matrices $\bm{B_l}$ and $\bm{B_{nl}}$ are linear and nonlinear geometrical matrices and $\vect{K}_S^{i j \tau s}=\vect{K}_0^{i j \tau s}+\vect{K}_{lnl}^{i j \tau s}+\vect{K}_{nll}^{i j \tau s}+\vect{K}_{nlnl}^{i j \tau s}$ is the FN of the secant stiffness matrix. In this paper, the nonlinear model considering the full components of the matrices $\bm{B_l}$ and $\bm{B_{nl}}$ for the Green–Lagrange strain vector is referred to as a Full NL model. The specific expressions of these matrices have been provided in \cite{wu2019large} and are omitted here for brevity.

In the case of conservative loading, the tangent stiffness matrix is obtained by linearizing the virtual variation of the internal strain energy, which is expressed as:
\begin{equation}\label{eq:Lint_T}
\delta (\delta L_\mathrm{int}) =
\,< \delta (\delta \vect{\epsilon}^{\textrm{T}} \vect{\sigma} ) >\, =\, < \delta \vect{\epsilon}^{\textrm{T}} \delta \vect{\sigma} > + < \delta (\delta \vect{\epsilon}^{\textrm{T}} ) \vect{\sigma} > \,= \delta \vect{q}_{\tau i}^{\textrm{T}} \vect{K}_{T}^{i j \tau s} \delta \vect{q}_{s j}
\end{equation}
where $\vect{K}_{T}^{i j \tau s}=\vect{K}_{0}^{i j \tau s} + \vect{K}_{T_1}^{i j \tau s} + \vect{K}_{\sigma}^{i j \tau s}$. 
The first term $< \delta \vect{\epsilon}^{\text{T}} \delta \vect{\sigma} >$ in Eq.~(\ref{eq:Lint_T}) requires the constitutive equation to be linearized. Therefore, we have:
\begin{equation}
\delta \vect{\sigma}= \delta ( \vect{C}\vect{\epsilon} ) = \vect{C} \delta \vect{\epsilon} = \vect{C}  (\vect{B}_l^{s j} + 2 \, \vect{B}_{nl}^{s j}) \delta \vect{q}_{s j}
\label{eq:hooke_lin}
\end{equation}
\begin{equation}\label{eq:Lint_T_1}
\renewcommand{\arraystretch}{2.0}
\begin{array}{lll}
< \delta \vect{\epsilon}^\textrm{T} \delta \vect{\sigma} > & = & \delta \vect{q}_{\tau i}^\textrm{T}  < (\vect{B}_l^{\tau i} + 2 \, \vect{B}_{nl}^{\tau i})^\textrm{T}  \vect{C} \,  (\vect{B}_l^{s j} + 2 \, \vect{B}_{nl}^{s j}) > \delta \vect{q}_{s j} \\
& = & \delta \vect{q}_{\tau i}^\textrm{T} \, \vect{K}_0^{i j \tau s} \delta \vect{q}_{sj} + \delta \vect{q}_{\tau i}^\textrm{T} \, \big( 2 \, \vect{K}_{lnl}^{i j \tau s} \big) \, \delta \vect{q}_{sj} + \delta \vect{q}_{\tau i}^\textrm{T} \, \vect{K}_{nll}^{i j \tau s}  \delta \vect{q}_{sj} + \delta \vect{q}_{\tau i}^\textrm{T} \, \big( 2 \, \vect{K}_{nlnl}^{i j \tau s} \big)  \delta \vect{q}_{sj} \\
& = & \delta \vect{q}_{\tau i}^\textrm{T} \, \big( \vect{K}_{0}^{i j \tau s} + \vect{K}_{T_1}^{i j \tau s} \big) \, \delta \vect{q}_{sj}
\end{array}
\end{equation}
where $\vect{K}_{T_1}^{i j \tau s}=2\,\vect{K}_{lnl}^{i j \tau s}+\vect{K}_{nll}^{i j \tau s}+2\,\vect{K}_{nlnl}^{i j \tau s}$ represents the nonlinear contribution of the FN of the tangent stiffness matrix resulting from the linearization of the constitutive relation. 

The evaluation of the second term $< \delta (\delta \vect{\epsilon}^\textrm{T} ) \vect{\sigma} >$ in Eq. (\ref{eq:Lint_T}) requires the linearization of the nonlinear geometrical relations. From Eqs. (\ref{eq1}) and (\ref{eq:bnl}) and according to Crisfield \cite{crisfiedlnl}, the linearization of the virtual variation of the strain vector reads as follows \cite{wu2019large}:
\begin{equation} \label{eq100}
\scriptsize{
\delta(\delta{\bm{\epsilon}}) = \left\{\begin{array}{c}
(\delta{u_{{x}_{,x}}})\textsubscript{v}\;\delta{u_{{x}_{,x}}} + (\delta{u_{{y}_{,x}}})\textsubscript{v}\;\delta{u_{{y}_{,x}}} + (\delta{u_{{z}_{,x}}})\textsubscript{v}\;\delta{u_{{z}_{,x}}}\\\\
(\delta{u_{{x}_{,y}}})\textsubscript{v}\;\delta{u_{{x}_{,y}}} + (\delta{u_{{y}_{,y}}})\textsubscript{v}\;\delta{u_{{y}_{,y}}} + (\delta{u_{{z}_{,y}}})\textsubscript{v}\;\delta{u_{{z}_{,y}}}\\\\
(\delta{u_{{x}_{,z}}})\textsubscript{v}\;\delta{u_{{x}_{,z}}} + (\delta{u_{{y}_{,z}}})\textsubscript{v}\;\delta{u_{{y}_{,z}}} + (\delta{u_{{z}_{,z}}})\textsubscript{v}\;\delta{u_{{z}_{,z}}}\\\\
\left[(\delta{u_{{x}_{,x}}})\textsubscript{v}\;\delta{u_{{x}_{,z}}} + \delta{u_{{x}_{,x}}}\;(\delta{u_{{x}_{,z}}})\textsubscript{v}\right] + \left[(\delta{u_{{y}_{,x}}})\textsubscript{v}\;\delta{u_{{y}_{,z}}} + \delta{u_{{y}_{,x}}}\;(\delta{u_{{y}_{,z}}})\textsubscript{v}\right] + \left[(\delta{u_{{z}_{,x}}})\textsubscript{v}\;\delta{u_{{z}_{,z}}} + \delta{u_{{z}_{,x}}}\;(\delta{u_{{z}_{,z}}})\textsubscript{v}\right]\\\\
\left[(\delta{u_{{x}_{,y}}})\textsubscript{v}\;\delta{u_{{x}_{,z}}} + \delta{u_{{x}_{,y}}}\;(\delta{u_{{x}_{,z}}})\textsubscript{v}\right] + \left[(\delta{u_{{y}_{,y}}})\textsubscript{v}\;\delta{u_{{y}_{,z}}} + \delta{u_{{y}_{,y}}}\;(\delta{u_{{y}_{,z}}})\textsubscript{v}\right] + \left[(\delta{u_{{z}_{,y}}})\textsubscript{v}\;\delta{u_{{z}_{,z}}} + \delta{u_{{z}_{,y}}}\;(\delta{u_{{z}_{,z}}})\textsubscript{v}\right]  \\\\
\left[(\delta{u_{{x}_{x}}})\textsubscript{v}\;\delta{u_{{x}_{,y}}} + \delta{u_{{x}_{,x}}}\;(\delta{u_{{x}_{,y}}})\textsubscript{v}\right] + \left[(\delta{u_{{y}_{,x}}})\textsubscript{v}\;\delta{u_{{y}_{,y}}} + \delta{u_{{y}_{,x}}}\;(\delta{u_{{y}_{,y}}})\textsubscript{v}\right] + \left[(\delta{u_{{z}_{,x}}})\textsubscript{v}\;\delta{u_{{z}_{,y}}} + \delta{u_{{z}_{,x}}}\;(\delta{u_{{z}_{,y}}})\textsubscript{v}\right] 
\end {array}\right\}
}
\end{equation} 
in which the subscript ``v" represents the variation. Thus, we have:
\begin{equation} \label{eq200}
\delta(\delta{\bm{\epsilon}}) = \vect{B}^{*}_{nl} \left\{\begin{array}{c}
\delta{q}_{x_{{\tau}i}}\delta{q}_{x_{sj}} \\\\
\delta{q}_{y_{{\tau}i}}\delta{q}_{y_{sj}} \\\\
\delta{q}_{z_{{\tau}i}}\delta{q}_{z_{sj}} 
\end {array}\right\}
\end{equation}
where
\begin{equation} \label{eq300}
\scriptsize{
\vect{B}^{*}_{nl} = \left[ \begin{array}{ccc}
F_{\tau}F_{s}N_{i_{,x}}N_{j_{,x}} & F_{\tau}F_{s}N_{i_{,x}}N_{j_{,x}} & F_{\tau}F_{s}N_{i_{,x}}N_{j_{,x}}\\\\
F_{\tau}F_{s}N_{i_{,y}}N_{j_{,y}} & F_{\tau}F_{s}N_{i_{,y}}N_{j_{,y}} & F_{\tau}F_{s}N_{i_{,y}}N_{j_{,y}} \\\\
F_{\tau_ {,z}}F_{s_{,z}}N_{i}N_{j} & F_{\tau_ {,z}}F_{s_{,z}}N_{i}N_{j} & F_{\tau_ {,z}}F_{s_{,z}}N_{i}N_{j} \\\\
F_{\tau}F_{s_{,z}}N_{i_{,x}}N_{j} + F_{\tau_{,z}}F_{s}N_{i}N_{j_{,x}} & F_{\tau}F_{s_{,z}}N_{i_{,x}}N_{j} + F_{\tau_{,z}}F_{s}N_{i}N_{j_{,x}} & F_{\tau}F_{s_{,z}}N_{i_{,x}}N_{j} + F_{\tau_{,z}}F_{s}N_{i}N_{j_{,x}} \\\\
F_{\tau_{,z}}F_{s}N_{i}N_{j_{,y}} + F_{\tau}F_{s_{,z}}N_{i_{,y}}N_{j} & F_{\tau_{,z}}F_{s}N_{i}N_{j_{,y}} + F_{\tau}F_{s_{,z}}N_{i_{,y}}N_{j} & F_{\tau_{,z}}F_{s}N_{i}N_{j_{,y}} + F_{\tau}F_{s_{,z}}N_{i_{,y}}N_{j} \\\\
F_{\tau}F_{s}N_{i_{,x}}N_{j_{,y}} + F_{\tau}F_{s}N_{i_{,y}}N_{j_{,x}} & F_{\tau}F_{s}N_{i_{,x}}N_{j_{,y}} + F_{\tau}F_{s}N_{i_{,y}}N_{j_{,x}} & F_{\tau}F_{s}N_{i_{,x}}N_{j_{,y}} + F_{\tau}F_{s}N_{i_{,y}}N_{j_{,x}} 
\end {array}\right]
}
\end{equation}

After simple mathematical manipulations, we obtain:
\begin{equation}\label{eq:Lint_sig}
\begin{array}{lll}
< \delta (\delta \vect{\epsilon}^\textrm{T} ) \vect{\sigma} > & = &
< \left\{\begin{array}{c}
\delta q_{u_{x_{\tau i}}} \delta q_{u_{x_{s j}}} \\\\
\delta q_{u_{y_{\tau i}}} \delta q_{u_{y_{s j}}} \\\\
\delta q_{u_{z_{\tau i}}} \delta q_{u_{z_{s j}}}
\end {array}\right\}^\textrm{T} (\vect{B}_{nl}^{*})^\textrm{T} \vect{\sigma} > \\\\
& = & \delta \vect{q}_{\tau i}^\textrm{T} < \mathrm{diag} \left[ (\vect{B}_{nl}^{*})^\textrm{T} \vect{\sigma} \right] > \delta \vect{q}_{s j} \\\\
& = & \delta \vect{q}_{\tau i}^\textrm{T} < \mathrm{diag} \left[ (\vect{B}_{nl}^{*})^\textrm{T} (\vect{\sigma}_l+\vect{\sigma}_{nl}) \right] > \delta \vect{q}_{s j} \\\\
& = & \delta \vect{q}_{\tau i}^\textrm{T} (\vect{K}_{\sigma_l}^{i j \tau s} + \vect{K}_{\sigma_{nl}}^{i j \tau s}) \delta \vect{q}_{s j} \\\\
& = & \delta \vect{q}_{\tau i}^\textrm{T} \vect{K}_{\sigma}^{i j \tau s} \delta \vect{q}_{s j}
\end{array}
\end{equation}
where the diagonal terms of the $3 \times 3$ diagonal matrix $\mathrm{diag} \left[ (\vect{B}_{nl}^{*})^\textrm{T} \vect{\sigma} \right]$ are the components of the vector $(\vect{B}_{nl}^{*})^\textrm{T} \vect{\sigma}$. According to Eqs. (\ref {eq1}) and (\ref{eq:hooke}), $\vect{\sigma}_l= \vect{C}\vect{\epsilon}_l$, $\vect{\sigma}_{nl}= \vect{C}\vect{\epsilon}_{nl}$. Furthermore, also the called \textit{geometric stiffness} matrix $\vect{K}_{\sigma}^{i j \tau s}=\vect{K}_{\sigma_l}^{i j \tau s} + \vect{K}_{\sigma_{nl}}^{i j \tau s}$ is defined, , which contributes to the tangent stiffness matrix arising from the strain–displacement geometrical relation. Its specific expression can be referred to the work \cite{wu2019large} and is omitted for simplicity. 

Once the FNs of secant and tangent stiffness matrices are available as the basic building blocks, one can expand them to formulate the nonlinear governing equations and incremental equations of the global system. Thus, the path-following Newton-Raphson linearization method (or tangent method) is chosen to compute the nonlinear system. 
Readers are referred to \cite{Pagani2018,wu2019large} for more information about the employed Newton-Raphson method with a path-following constraint and the explicit forms of tangent and secant stiffness matrices. 
\section{Numerical Results} \label{V}
This section presents numerical results of the large deflection and post-buckling of composite plates based on CUF plate models. First, different refined composite plates under uniform transverse pressure are calculated with symmetric and antisymmetric laminations. The equilibrium curves are compared with those in the available literature. Then, the post-buckling analyses of laminated plates subjected to in-plane compressive loads are performed, and a comprehensive investigation is carried out for the evaluation of the effects of different edge conditions on the plate post-buckling behaviours.
\subsection{Large deflection of composite plates subjected to uniform transverse pressure} 
%
\subsubsection{Cross-ply [0/90]$_{s}$ laminate with different edge conditions}
For the first analysis case, a 4-layer $[0/90]_{s}$ square composite plate is studied. The geometric characteristics of the structure are with width $a=b=30.48$~cm and thickness $h=7.62$~mm. This structure is subjected to a uniform transverse pressure. The transverse pressure is fixed in the direction ($z$ axis), and the pressure values are investigated versus the corresponding displacements along the equilibrium path. The following two kinds of boundary conditions are considered for this case: (a) all edges are fully clamped in such a way that $u=v=w=0$ at $x=0,a$ and $y=0,b$; (b) all edges are simply-supported in such a way that $u=v=w=0$ at $x=0,a,\:z=0$ and $y=0,b,\:z=0$. The material properties for this composite plate are reported in Table~\ref{ex2properties}.
%
%
%
\begin{table}[htbp]
\centering
\begin{tabular}{cccccc}
\toprule
\toprule 
E$_{1}$ (GPa) & $\text{E}_{2}=\text{E}_{3}$ (GPa) & $\text{G}_{12}=\text{G}_{13}$ (GPa) & $\nu_{12}=\nu_{13}$ \\
\midrule
12.60 &  12.62 & 2.15 &  0.2395 \\
\bottomrule
\bottomrule
\end{tabular} 
\caption{Material properties of a 4-layer $[0/90]_{s}$ composite plate \cite{reddy2003mechanics}.}\label{ex2properties}
\end{table}

\begin{figure}[htbp]
\centering
\subfloat[Mesh approximation\label{ex2a-convergexy}]{
\includegraphics[width=0.44\textwidth]{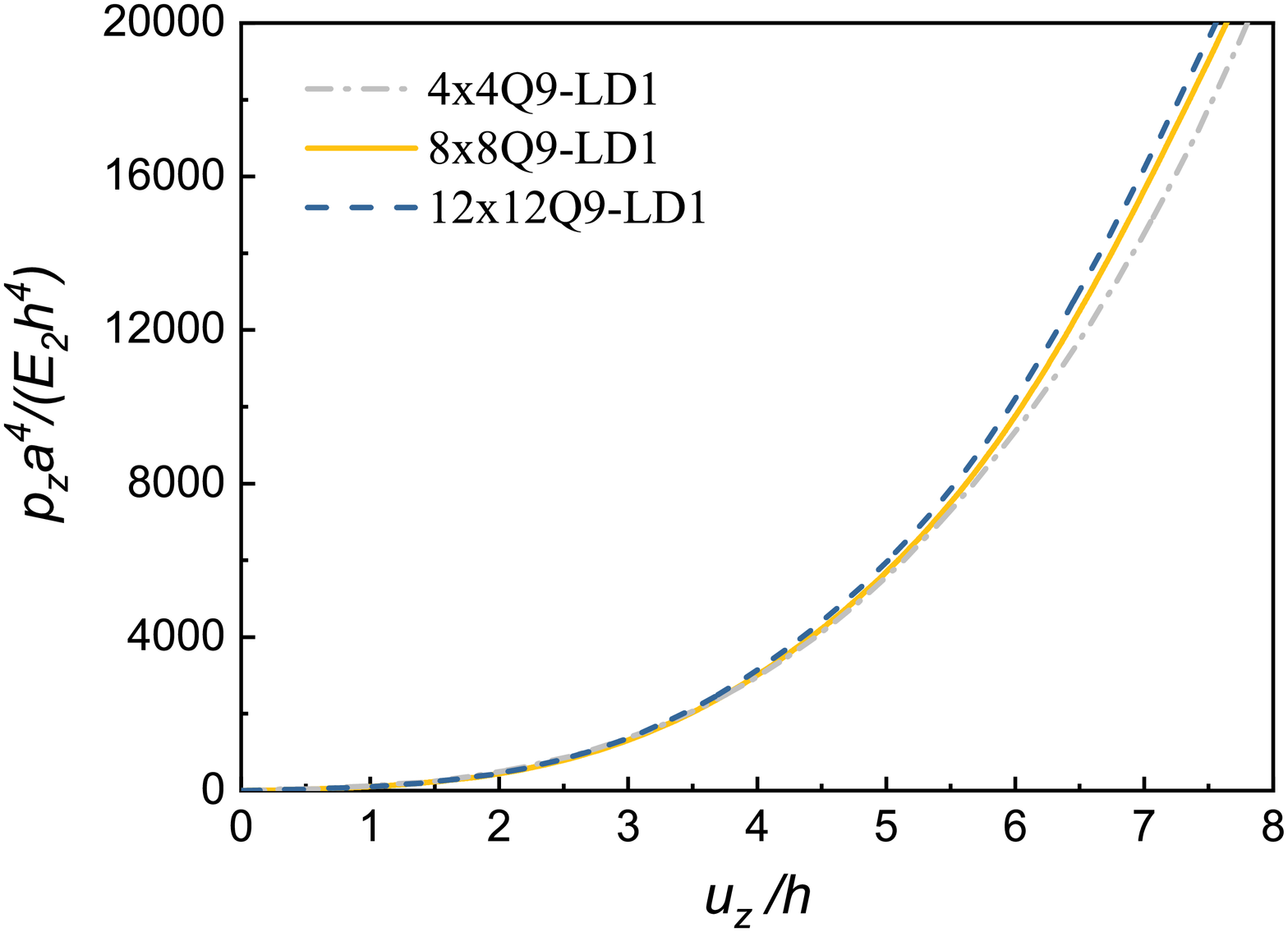}}\hfill
\subfloat[Kinematic expansion\label{ex2a-convergez}]{
\includegraphics[width=0.44\textwidth]{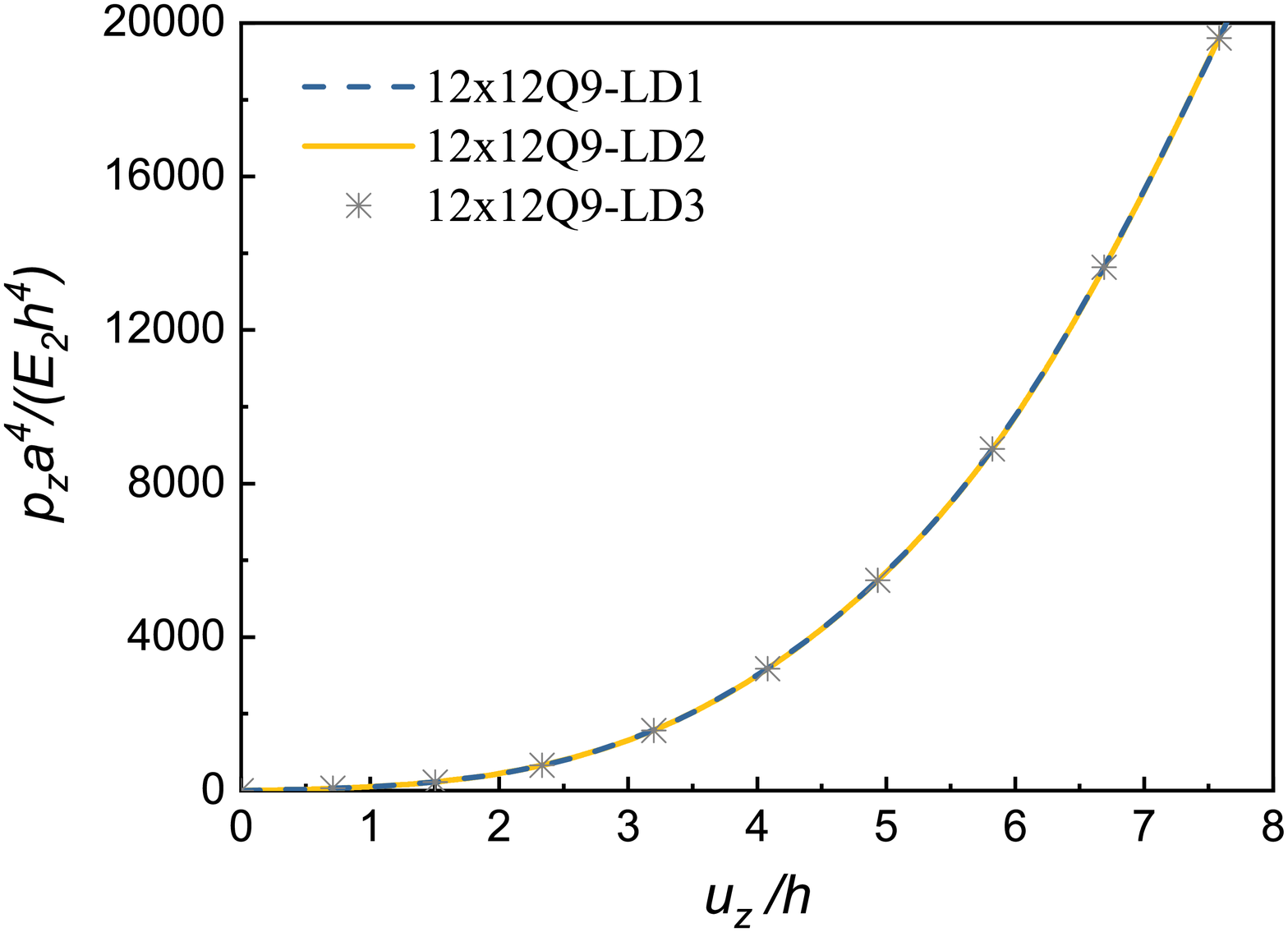}}\hfill
\caption{Convergence analysis for a 4-layer $[0/90]_{s}$ composite plate under uniform transverse pressure with clamped edge conditions at the center of the composite plate. Comparison of (a) various in-plane mesh approximations and (b) different orders of Lagrange expansion functions in the thickness direction.} 
\label{ex2a-converge}
\end{figure}

In this work, convergence analyses illustrated in Fig.~\ref{ex2a-converge} are conducted to evaluate the effects of mesh approximation and kinematic expansion.  First, the finite plate elements are 4$\times$4Q9, 8$\times$8Q9, and 12$\times$12Q9 with the fixed LD1 theory approximation order for each layer. Then, the kinematic expansion order along the thickness direction is changed from LD1 to LD3, considering the 12$\times$12Q9 in-plane mesh approximation. 
Moreover, the transverse displacement values for various models and loads, along with the total degrees of freedom (DOFs), are reported in Table \ref{Table001}. 
\begin{table}[htbp]
\centering
\begin{tabular}{cccccc}
\toprule
\toprule 
\multirow{2}{*}{CUF plate model}  & \multirow{2}{*}{DOFs}  &   \multicolumn{2}{c}{$u_{z}/h$ } \\
\cmidrule(l){3-4} 
 &  & $p_{z}a^4/E_{2}h^4=4000$ & $p_{z}a^4/E_{2}h^4=16000$ \\
\midrule
4  $\times$ 4Q9-LD1  &  1215  & 4.44 &  7.16 \\
8  $\times$ 8Q9-LD1  &  4335  & 4.42 &  7.03 \\
12 $\times$ 12Q9-LD1  &  9375  & 4.41 &  6.98 \\
\bottomrule
\bottomrule
\end{tabular} 
\caption{Equilibrium points of nonlinear response curves of a 4-layer $[0/90]_{s}$ composite plate under transverse pressure with clamped edge conditions for different models and loads at the center of the composite plate.}\label{Table001}
\end{table}
As observed from Fig. \ref{ex2a-converge} and Table \ref{Table001}, the convergence is achieved for the nonlinear response curves based on the 12$\times$12Q9-LD1 model, which will be used to investigate the equilibrium curves of the above-mentioned composite plate. Furthermore, the results show that the difference between the equilibrium paths for the investigated CUF plate models is not significant in the case of the composite plate under bending.

\begin{figure}[htbp]
\centering
\subfloat[Clamped edge conditions\label{ex2a}]{
\includegraphics[width=0.45\textwidth]{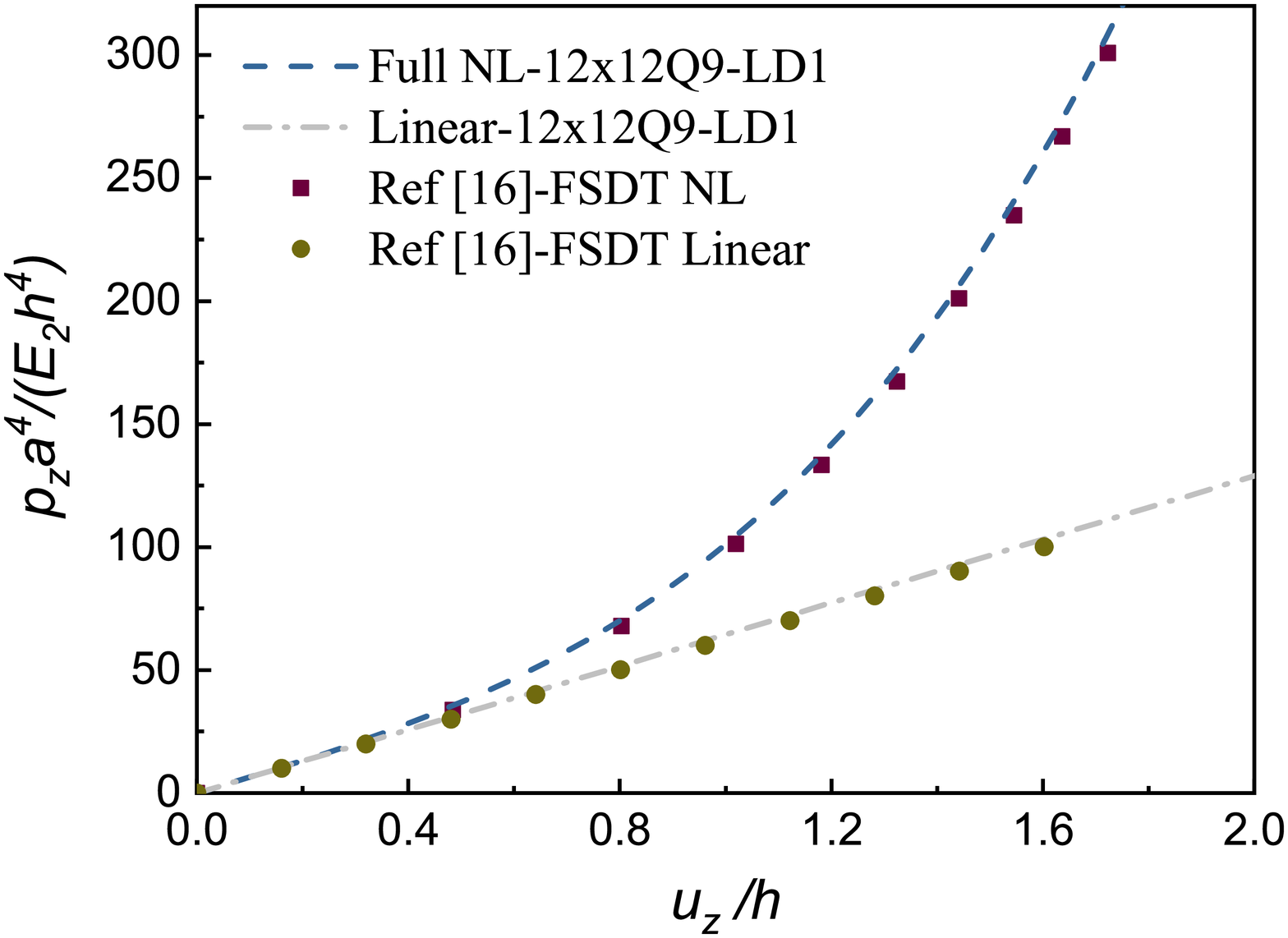}}\hfill
\subfloat[Simply-supported edge conditions\label{ex2b}]{
\includegraphics[width=0.45\textwidth]{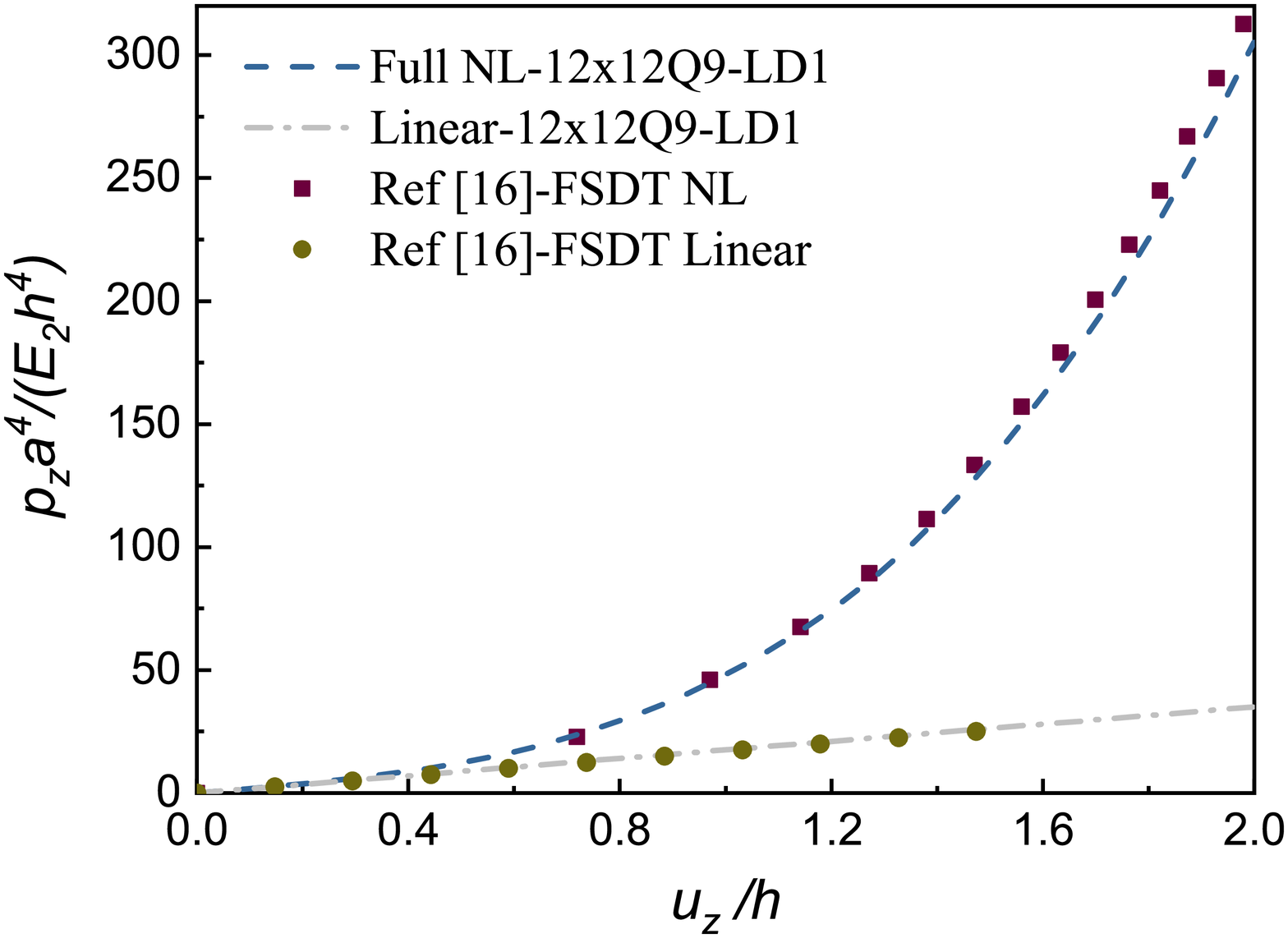}}\hfill
\caption{Comparison of the equilibrium curves for a 4-layer $[0/90]_{s}$ composite plate under uniform transverse pressure with different edge conditions.} 
\label{ex2}
\end{figure}

The equilibrium curves for this composite plate subjected to both clamped and simply-supported edge conditions are illustrated in Fig.~\ref{ex2}, which plots the normalized values of the displacement in the middle point of the plate versus the normalized values of the applied transverse pressure. As shown in this figure, the equilibrium curves predicted by the CUF linear and Full nonlinear (Full NL) plate models match well with those in the available literature using the FSDT theory \cite{reddy2003mechanics}. In addition, the difference between linear and nonlinear models is more significant as the transverse pressure value is increased. Also, the load-carrying capacity of the composite plate with the clamped edge conditions is higher than that of the composite plate with the simply-supported ones. 

Table~\ref{ex2xytable} shows the displacement values based on different 2D CUF models and solutions in the available literature \cite{reddy2003mechanics} at the fixed load of $\frac{{{P_z}{a^4}}}{{E_{2}{h^4}}} = 100$ for the clamped edge conditions, and at the fixed load of $\frac{{{P_z}{a^4}}}{{E_{2}{h^4}}} = 25$ for the simply-supported edge conditions. According to this table, the displacement values of the 2D CUF Full NL and linear models agree well with the corresponding values of the FSDT nonlinear and linear models, respectively.
%
%
%
\begin{table}[htbp]
\centering
\begin{tabular}{cccccc}
\toprule
\toprule 
\multirow{2}{*}{Model}   &   \multicolumn{1}{c}{Clamped} & \multicolumn{1}{c}{Simply-supported} \\
\cmidrule(l){2-2} \cmidrule(l){3-3} 
 & $u_{z} (mm)$ & $u_{z} (mm)$\\
\midrule
Full NL 12$\times$12Q9-LD1 &  7.57 & 5.62 \\
Ref \cite{reddy2003mechanics} - FSDT NL &  7.71 & 5.67 \\
Linear 12$\times$12Q9-LD1 &  11.81 & 10.86 \\
Ref \cite{reddy2003mechanics} - FSDT Linear &  12.19 & 11.21 \\
\bottomrule
\bottomrule
\end{tabular} 
\caption{The displacement values based on different 2D CUF models and solutions in the available literature \cite{reddy2003mechanics} for the 4-layer $[0/90]_{s}$ composite plates under transverse pressure at the fixed load of $\frac{{{P_z}{a^4}}}{{E_{2}{h^4}}} = 100$ with clamped edge conditions, and at the fixed load of $\frac{{{P_z}{a^4}}}{{E_{2}{h^4}}} = 25$ with simply-supported edge conditions.}\label{ex2xytable}
\end{table}
\subsubsection{[45/-45/0/0/45/-45/90/90]$_{s}$ laminate with clamped edge conditions}
\begin{figure}[htbp]
\centering
\includegraphics[width=0.6\textwidth,angle=0]{./
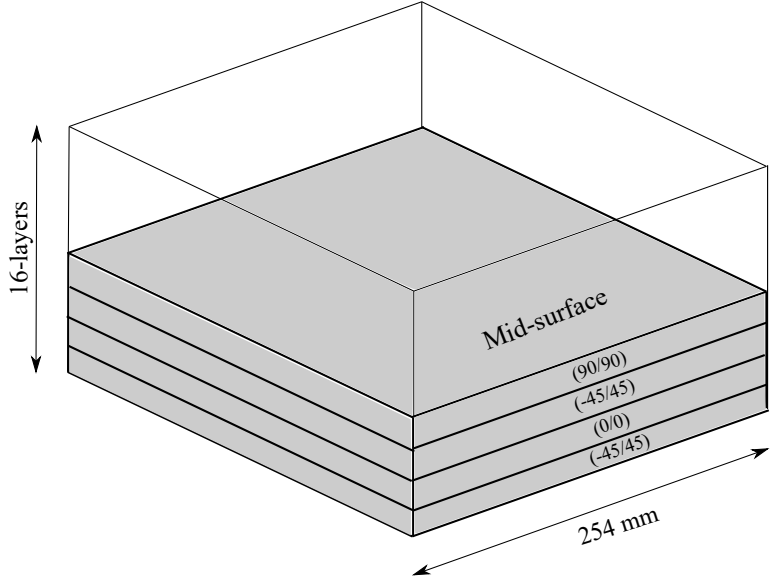} 
\hspace{0.00\textwidth}
\caption{Schematic view of a 16-layer $[45/-45/0/0/45/-45/90/90]_{s}$ composite plate.}
\label{ex3schematic}
\end{figure}

A 16-layer $[45/-45/0/0/45/-45/90/90]_{s}$ square composite plate is analyzed as the second case. The geometric characteristics are with width $a=b=25.4$~cm and thickness $h=2.11$~mm. A schematic view of the investigated composite plate is illustrated in Fig.~\ref{ex3schematic}. The plate generates large deflection due to a uniform transverse pressure, and the edges are fully clamped so that $u=v=w=0$ at $x=0,a$ and $y=0,b$. The material properties for this composite plate are reported in Table~\ref{ex3properties}.  
%
%
\begin{table}[htbp]
\centering
\begin{tabular}{cccccc}
\toprule
\toprule 
$\text{E}_{1}$ (GPa) & $\text{E}_{2}=\text{E}_{3}$ (GPa) & $\text{G}_{12}=\text{G}_{13}$ (GPa) & $\text{G}_{23}$ (GPa) & $\nu_{12}=\nu_{13}$ \\
\midrule
131 &  13.03 & 6.41 &  4.72 &  0.38 \\
\bottomrule
\bottomrule
\end{tabular} 
\caption{Material properties of a 16-layer $[45/-45/0/0/45/-45/90/90]_{s}$ composite plate \cite{han2006postbuckling}.}\label{ex3properties}
\end{table}

\begin{figure}[htbp]
\centering
\subfloat[Mesh approximation\label{ex3-convergexy}]{
\includegraphics[width=0.44\textwidth]{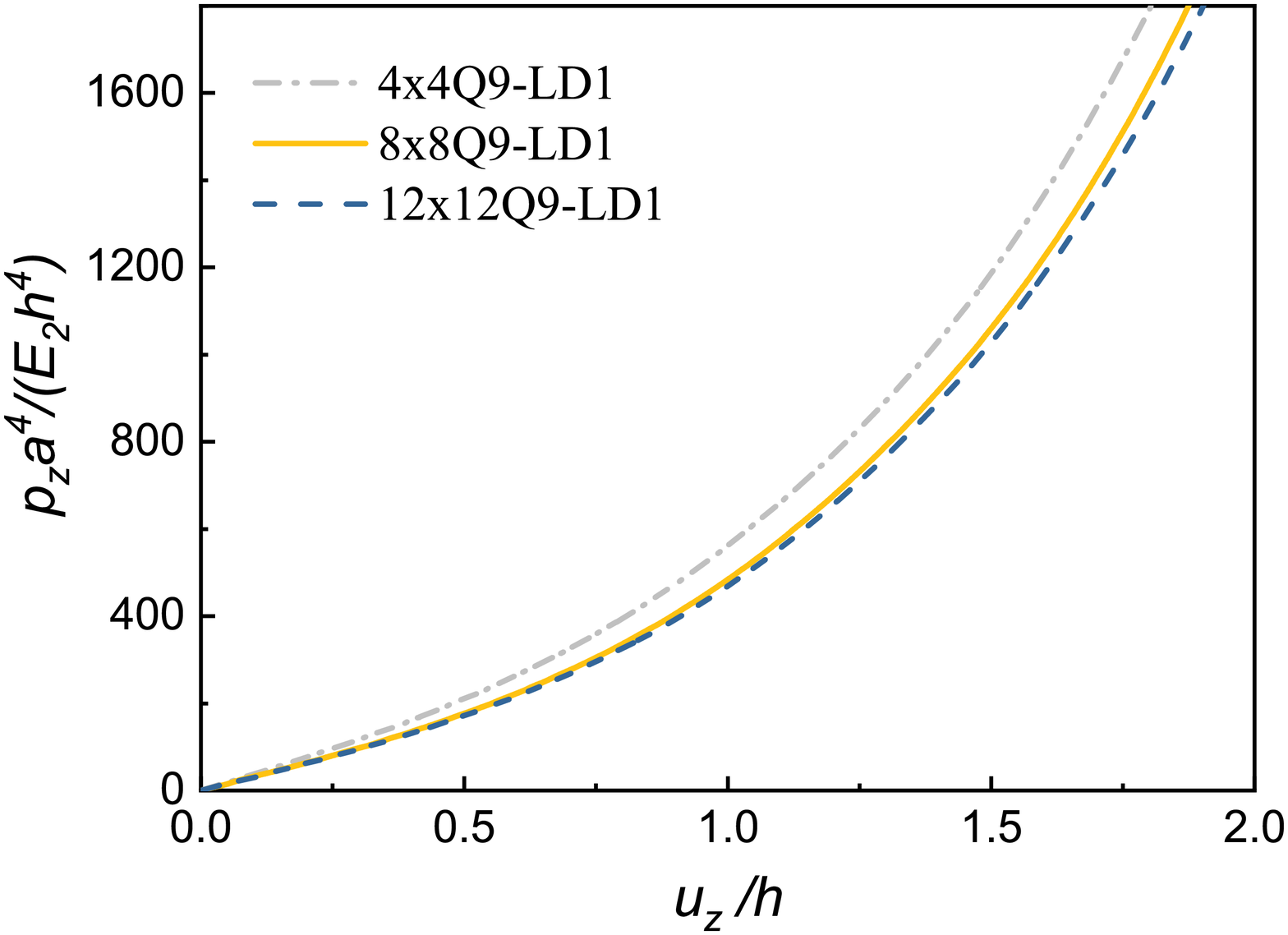}}\hfill
\subfloat[Kinematic expansion\label{ex3-convergez}]{
\includegraphics[width=0.44\textwidth]{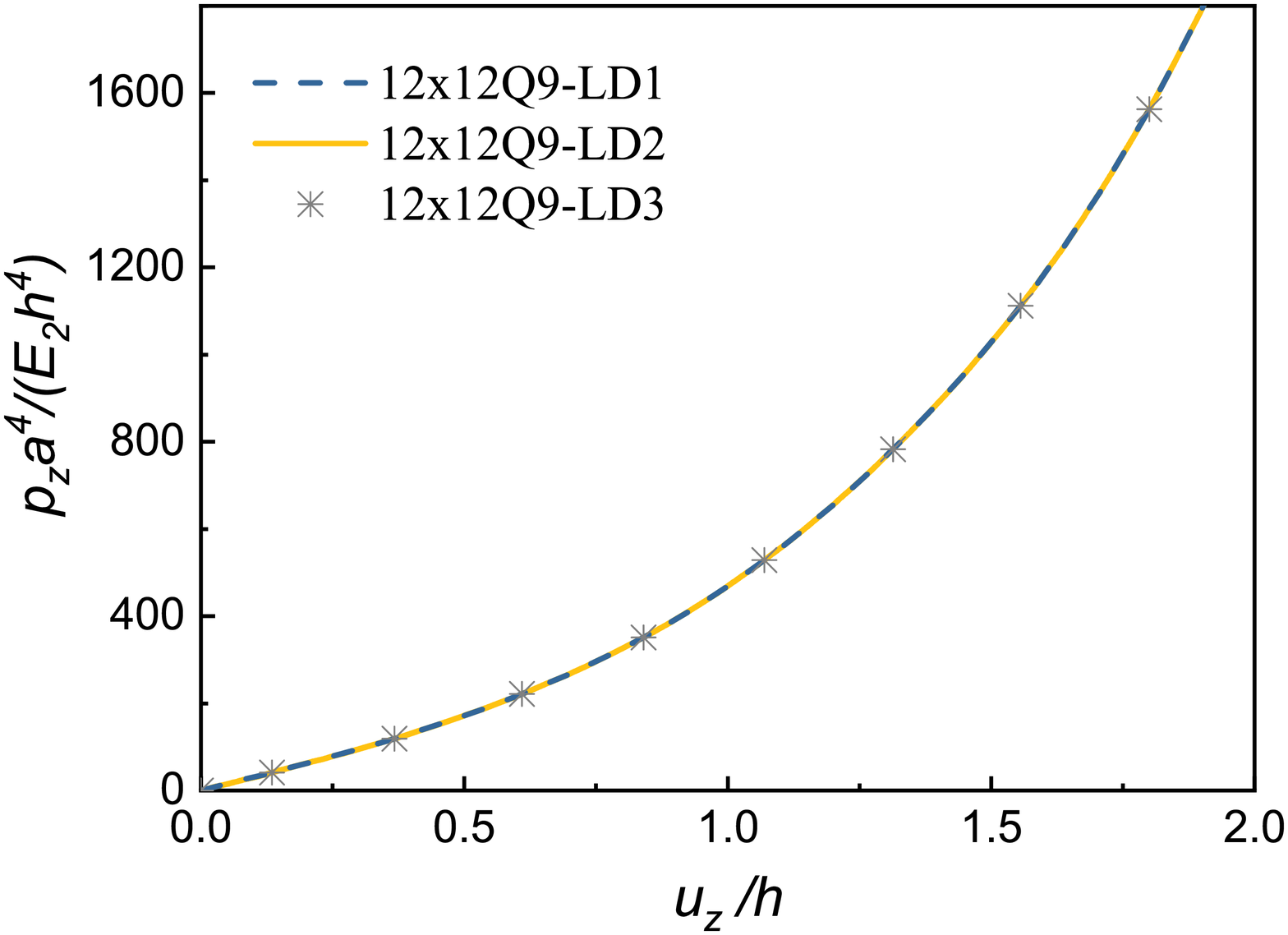}}\hfill
\caption{Convergence analysis for a 16-layer $[45/-45/0/0/45/-45/90/90]_{s}$ composite plate under uniform transverse pressure with clamped edge conditions.} 
\label{ex3-converge}
\end{figure}

The convergence analysis for this laminated case is illustrated in Fig.~\ref{ex3-converge}. 
Fig.~\ref{ex3-converge}a provides the transverse deflection at the center of the composite plate for various CUF plate models, and the in-plane meshes from 16Q9 to 144Q9 FEs are used, whereas only one LD1 is adopted for each layer in the thickness direction. Instead, analyses based on different through-the-thickness kinematic approximations are reported in Fig.~\ref{ex3-converge}b. 
Moreover, the transverse displacement values for various CUF plate models and loads are tabulated in Table \ref{Table002} along with the DOFs. 
\begin{table}[htbp]
\centering
\begin{tabular}{cccccc}
\toprule
\toprule 
\multirow{2}{*}{CUF plate model}  & \multirow{2}{*}{DOFs}  &   \multicolumn{2}{c}{$u_{z}/h$ } \\
\cmidrule(l){3-4} 
 &  & $p_{z}a^4/E_{2}h^4=400$ & $p_{z}a^4/E_{2}h^4=1600$ \\
\midrule
4  $\times$ 4Q9-LD1  &  4131  & 0.81 &  1.71 \\
8  $\times$ 8Q9-LD1  &  14739  & 0.89 &  1.80 \\
12 $\times$ 12Q9-LD1  &  31875  & 0.90 &  1.82 \\
\bottomrule
\bottomrule
\end{tabular} 
\caption{Equilibrium points of nonlinear response curves of a 16-layer $[45/-45/0/0/45/-45/90/90]_{s}$ composite plate under uniform transverse pressure with clamped edge conditions for different CUF plate models and loads. The displacement is calculated at the center of the composite plate.}\label{Table002}
\end{table}
As shown in Fig. \ref{ex3-converge} and Table \ref{Table002}, the convergence is obtained for the nonlinear static response when adopting the 12$\times$12Q9-LD1 model. 

\begin{figure}[htbp]
\centering
\includegraphics[width=0.6\textwidth,angle=0]{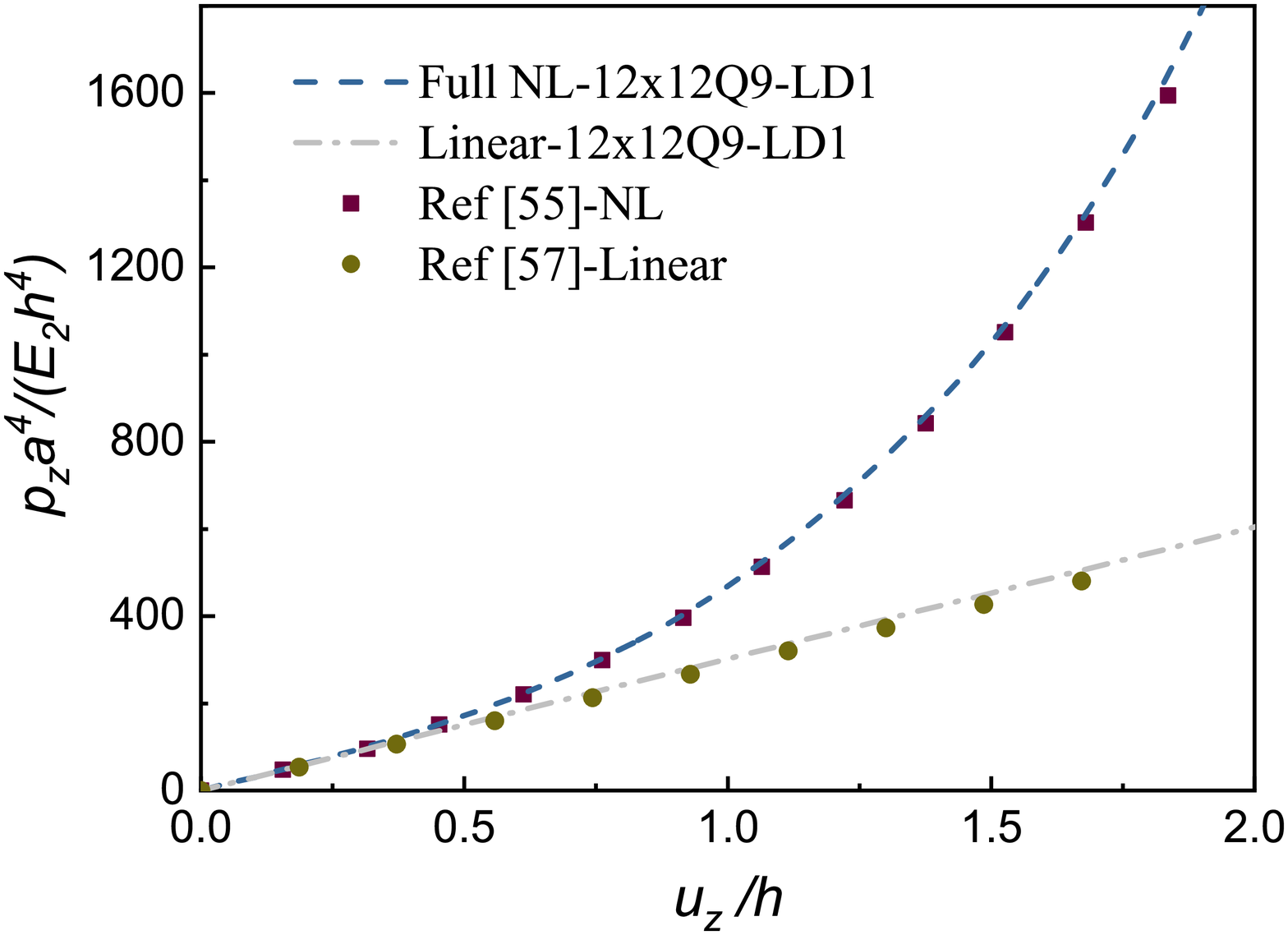}
\hspace{0.00\textwidth}
\caption{Comparison of the equilibrium curves for a 16-layer $[45/-45/0/0/45/-45/90/90]_{s}$ composite plate under uniform transverse pressure with clamped edge conditions.}
\label{ex3}
\end{figure}

Figure~\ref{ex3} depicts the equilibrium curves at the center of the laminated plate and the comparison with reference solutions in the available literature. As illustrated in this figure, the equilibrium curves obtained by the CUF linear and Full NL plate models agree well with the corresponding values from the available literature \cite{han2006postbuckling,lee1998nine,noor1976anisotropy}. 
\subsubsection{Cross-ply [0/90] and [0/90]$_{3}$ laminates with clamped edge conditions}
The last large-deflection cases are 2-layer $[0/90]$, and 6-layer $[0/90]_{3}$ square composite plates with the width of $a=b=30.48$~cm and the thickness of $h=2.44$~mm. The plates are subjected to the large deflection due to a uniform transverse pressure, and the edges are fully clamped such that $u=v=w=0$ at $x=0,a$ and $y=0,b$. The material properties for the two composite plates are shown in Table~\ref{ex1properties}.
%
%
\begin{table}[htbp]
\centering
\begin{tabular}{cccccc}
\toprule
\toprule 
$\text{E}_{1}$ (GPa) & $\text{E}_{2}=\text{E}_{3}$ (GPa) & $\text{G}_{12}=\text{G}_{13}$ (GPa) & $\text{G}_{23}$ (GPa) & $\nu_{12}=\nu_{13}$ \\
\midrule
275.79 &  6.89 & 4.13 &  3.44 &  0.25 \\
\bottomrule
\bottomrule
\end{tabular} 
\caption{Material properties of a 2-layer $[0/90]$ composite plate \cite{reddy2003mechanics}.}\label{ex1properties}
\end{table}

\begin{figure}[htbp]
\centering
\subfloat[Mesh approximation\label{ex1a-convergexy}]{
\includegraphics[width=0.48\textwidth]{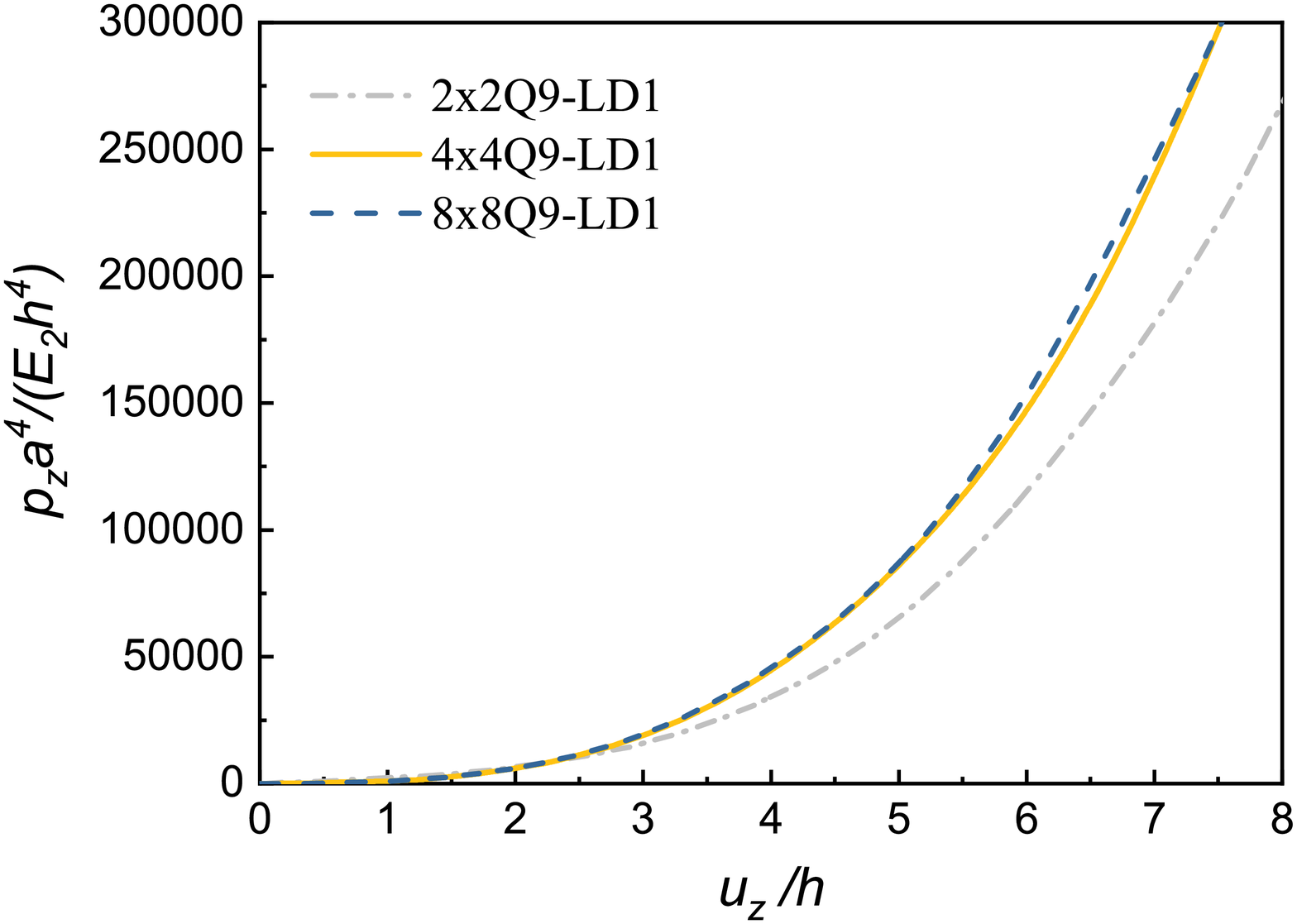}}\hfill
\subfloat[Kinematic expansion\label{ex1a-convergez}]{
\includegraphics[width=0.48\textwidth]{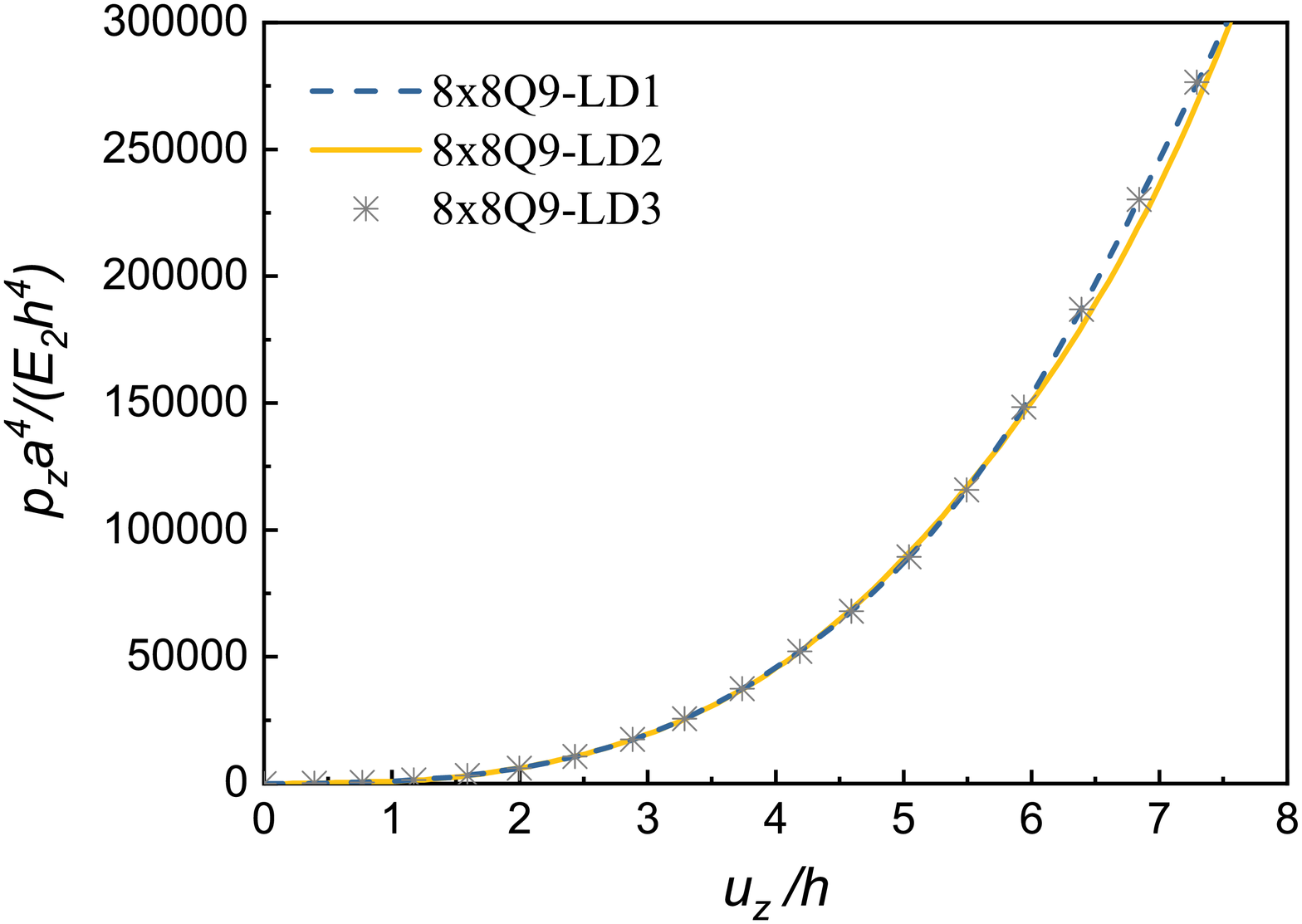}}\hfill
\caption{Convergence analysis for a 2-layer $[0/90]$ composite plate under uniform transverse pressure with clamped edge conditions.} 
\label{ex1-converge}
\end{figure}

In this case, the convergence analysis shown in Fig.~\ref{ex1-converge} is conducted to evaluate the effects of in-plane mesh kinematic expansion approximations. First, the finite plate elements are 2$\times$2Q9, 4$\times$4Q9, and 8$\times$8Q9 with the fixed LD1 kinematic approximation order for each layer. Then, the theoretical expansion order along the thickness direction is changed from LD1 to LD3, while the finite plate elements are fixed at 8$\times$8Q9. 
Moreover, the transverse displacement values for various CUF plate models and loads, along with the DOFs, are provided in Table \ref{Table003}. 
As can be seen in Fig.~\ref{ex1-converge}a, the convergence is achieved in the upper bound and at least for the 4$\times$4Q9 plate model, while the models with different through-the-thickness kinematic approximations are converged in Fig.~\ref{ex1-converge}b. 
Therefore, it is clear from Fig.~\ref{ex1-converge} and Table \ref{Table003} that the convergence can be reached for the nonlinear response when adopting the 8$\times$8Q9-LD1 CUF plate model. 
\begin{table}[htbp]
\centering
\begin{tabular}{cccccc}
\toprule
\toprule 
\multirow{2}{*}{CUF plate model}  & \multirow{2}{*}{DOFs}  &   \multicolumn{2}{c}{$u_{z}/h$ } \\
\cmidrule(l){3-4} 
 &  & $p_{z}a^4/E_{2}h^4=50000$ & $p_{z}a^4/E_{2}h^4=250000$ \\
\midrule
2  $\times$ 2Q9-LD1  & 225   & 4.61 & 7.80  \\
4  $\times$ 4Q9-LD1  &  729  & 4.18 & 7.07 \\
8  $\times$ 8Q9-LD1  &  2601  & 4.17 & 7.05  \\
\bottomrule
\bottomrule
\end{tabular} 
\caption{Equilibrium points of nonlinear response curves of a 2-layer $[0/90]$ composite plate under uniform transverse pressure with clamped edge conditions for different CUF plate models and loads. The displacement is calculated at the center of the plate structure.}\label{Table003}
\end{table}

\begin{figure}[htbp]
\centering
\subfloat[2-layer \label{ex1a}]{
\includegraphics[width=0.48\textwidth]{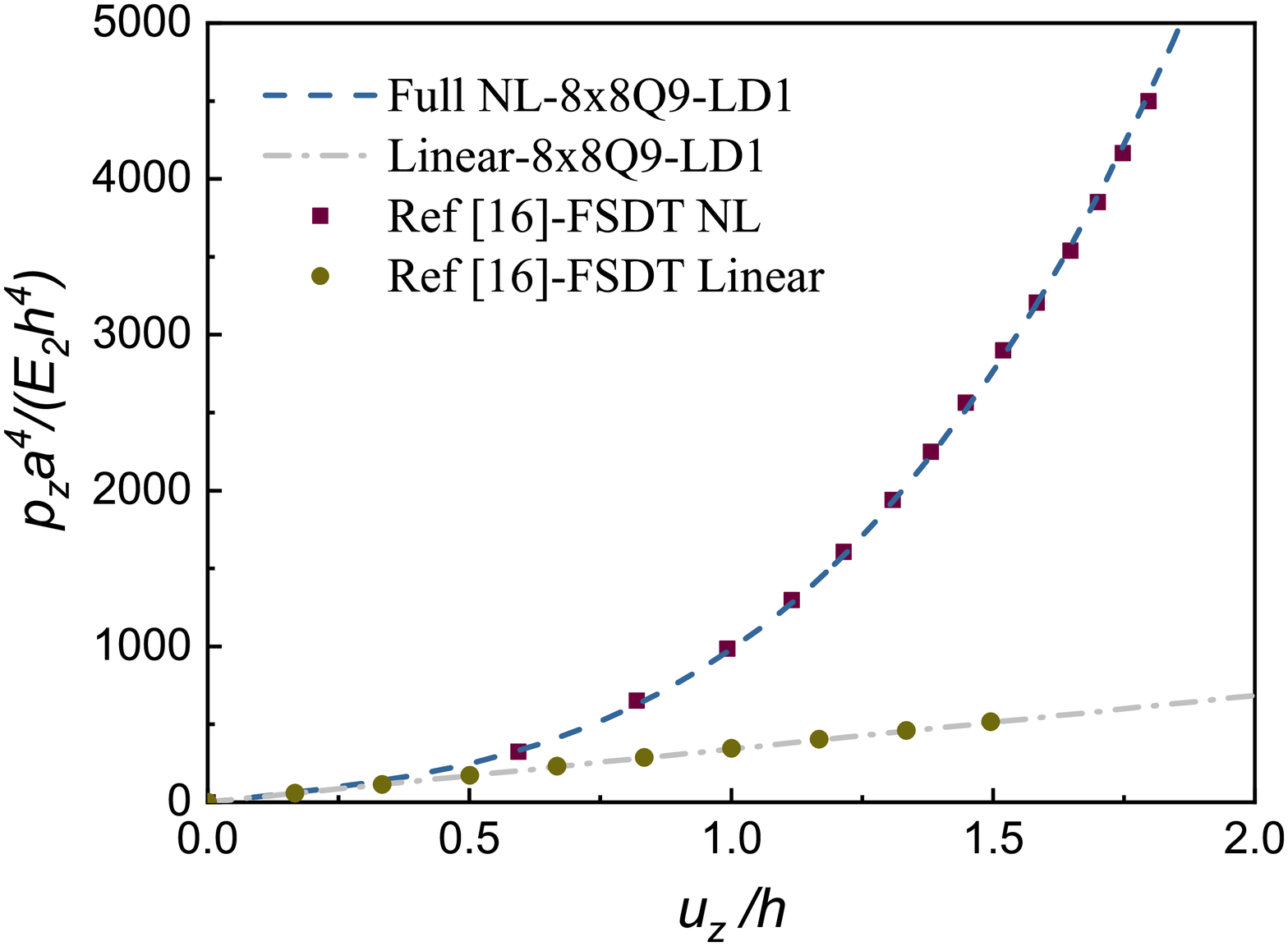}}\hfill
\subfloat[6-layer \label{ex1b}]{
\includegraphics[width=0.48\textwidth]{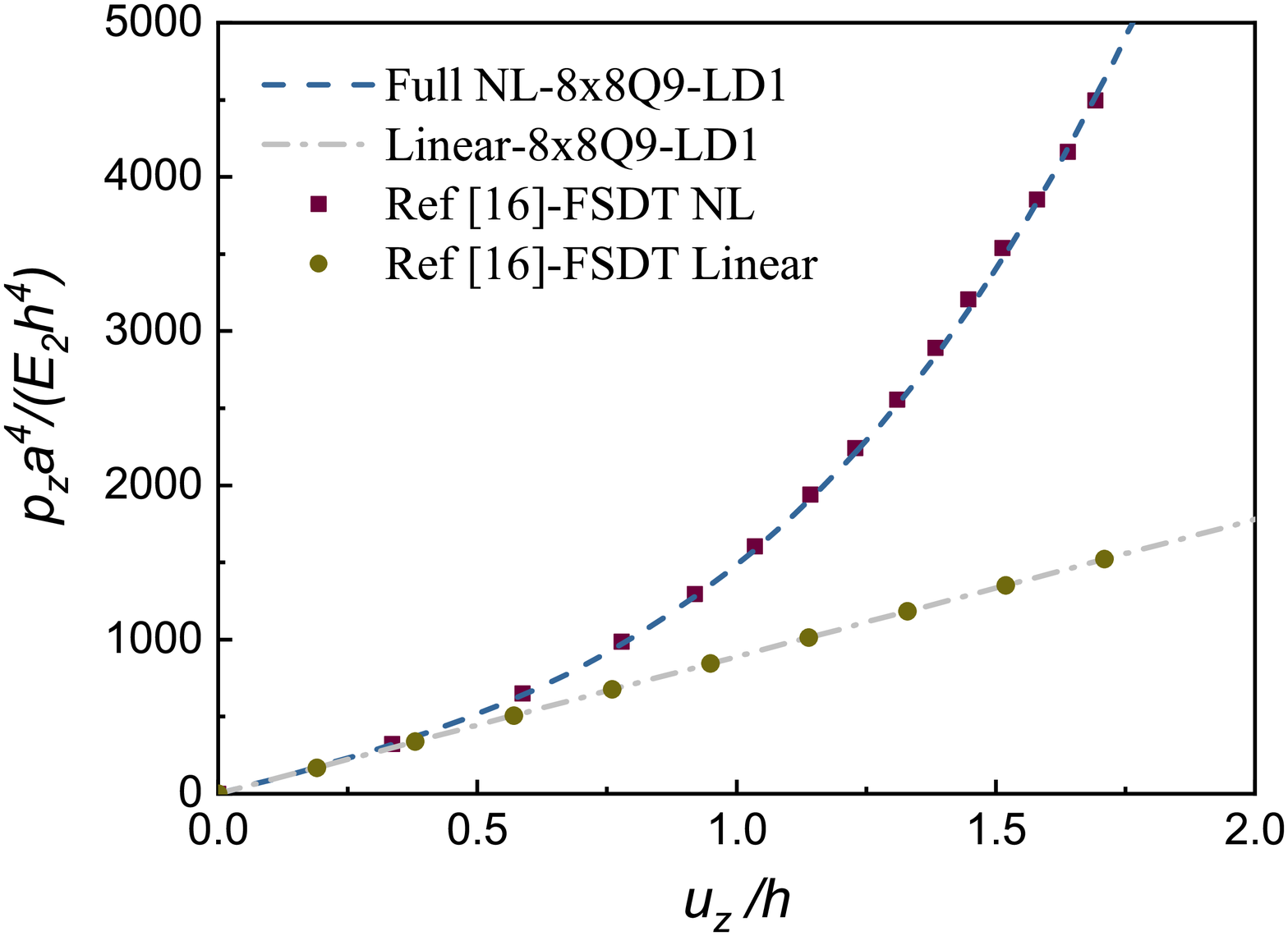}}\hfill
\caption{Comparison of the equilibrium curves for 2-layer [0/90] and 6-layer [0/90]$_{3}$ composite plates under uniform transverse pressure with clamped edge conditions.} 
\label{ex1}
\end{figure}

The equilibrium curves for 2-layer $[0/90]$ and 6-layer $[0/90]_3$ composite plates subjected to clamped edge conditions are shown in Fig.~\ref{ex1}, which plots the normalized values of the displacement at the center of the plate versus the normalized values of the applied transverse pressure. It is evident in this figure that the equilibrium curves obtained by the CUF linear and Full NL plate models provide excellent predictions compared with the solutions in the available literature using the FSDT theory \cite{reddy2003mechanics}. In addition, the load-carrying capacity of the composite plate with a 6-layers is higher than that of the composite plate with a 2-layers.

Table~\ref{ex2xytable} shows the displacement values based on the different 2D CUF plate models and solutions in the available literature \cite{reddy2003mechanics} at the fixed load of $\frac{{{P_z}{a^4}}}{{E_{2}{h^4}}} = 500$ for the 2-layer [0/90] composite plate, and at the fixed load of $\frac{{{P_z}{a^4}}}{{E_{2}{h^4}}} = 1500$ for the 6-layer [0/90/0/90/0/90] composite plate. Based on Table~\ref{ex2xytable}, the displacement values of the CUF linear and Full NL plate models match well with those of the FSDT nonlinear and linear models, respectively.
%
%
\begin{table}[htbp]
\centering
\begin{tabular}{cccccc}
\toprule
\toprule 
\multirow{2}{*}{Model}   &   \multicolumn{1}{c}{2-layer [0/90]} & \multicolumn{1}{c}{6-layer [0/90/0/90/0/90]} \\
\cmidrule(l){2-2} \cmidrule(l){3-3} 
 & $u_{z} (mm)$ & $u_{z} (mm)$\\
\midrule
Full NL 8$\times$8Q9-LD1 &  1.79 & 2.45 \\
Ref \cite{reddy2003mechanics} - FSDT NL &  1.78 & 2.43 \\
Linear 8$\times$8Q9-LD1 &  3.57 & 4.12 \\
Ref \cite{reddy2003mechanics} - FSDT Linear &  3.52 & 4.11 \\
\bottomrule
\bottomrule
\end{tabular} 
\caption{The displacement values based on different 2D CUF plate models and solutions in the available literature \cite{reddy2003mechanics} at the fixed load of $\frac{{{P_z}{a^4}}}{{E_{2}{h^4}}} = 500$ for the 2-layer [0/90] composite plate, and at the fixed load of $\frac{{{P_z}{a^4}}}{{E_{2}{h^4}}} = 1500$ for the 6-layer [0/90]$_{3}$ composite plate under uniform transverse pressure.}\label{ex1xytable}
\end{table}
\subsection{Post-buckling of composite plates under in-plane compressive loads} 
%
\subsubsection{Cross-ply [0/90]$_{2}$ rectangular laminate with simply-supported edge conditions}
\begin{figure}[htbp]
\centering
\includegraphics[width=0.65\textwidth,angle=0]{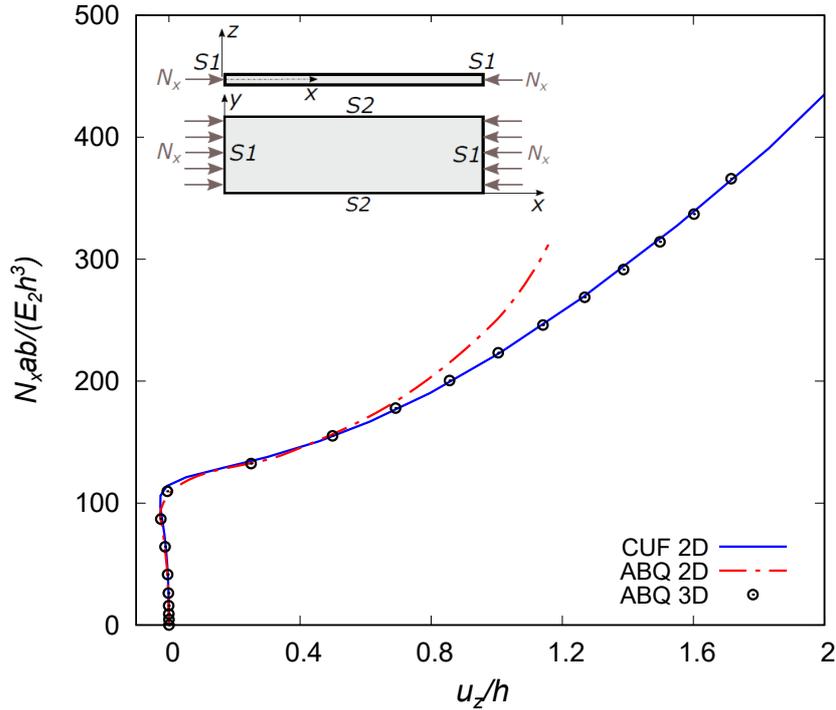}
\hspace{0.00\textwidth}
\caption{Comparison of equilibrium curves for a cross-ply $[0/90]_{2}$ laminate under in-plane compressive line loads in the $x$-axis direction with simply-supported edge conditions based on the 2D CUF Full NL model (20$\times$5Q9-LD1), ABQ 2D NL model (60$\times$15 S8R) and ABQ 3D NL model (60$\times$15$\times$4 C3D20R).}
\label{ex10a}
\end{figure}

A 4-layer $[0/90]_{2}$ rectangular composite plate is considered as the first post-buckling case. The structure has the length of $a=20$~cm, the width of $b=5$~cm, and the thickness of $h=2$~mm. The plate is subjected to in-plane compressive line loads in the $x$-axis direction, $N_{x}$ (force per unit width), see Fig. \ref{ex10a}. 
The edges are simply-supported in such a way that one set of opposite edges along width $x=0,a$ satisfy $v=w=0$ (see S1 in Fig.~\ref{ex10a}), whereas another set of simply-supported opposite edges along the length $y=0,b$ satisfy $w=0$ at $z=0$ (see S2 in Fig.~\ref{ex10a}). Furthermore, a constraint condition satisfying $u=v=0$ at the center point of the plate is used in order to avoid the rigid-body motion of the plate. The material properties, loading, and edge conditions for this composite plate are shown in Table~\ref{ex10properties} and Fig.~\ref{ex10a}. 
%
%
\begin{table}[htbp]
\centering
\begin{tabular}{cccccc}
\toprule
\toprule 
E$_{1}$ (GPa) & $\text{E}_{2}=\text{E}_{3}$ (GPa) & $\text{G}_{12}=\text{G}_{13}$ (GPa) & G$_{23}$ (GPa) & $\nu_{12}=\nu_{13}$ \\
\midrule
220 &  5.5 & 3.3 &  2.75 &  0.25 \\
\bottomrule
\bottomrule
\end{tabular} 
\caption{Material properties of a 4-layer $[0/90]_{2}$ composite plate.}\label{ex10properties}
\end{table}

\begin{figure}[htbp]
\centering
\subfloat[Mesh approximation\label{ex10-convergexy}]{
\includegraphics[width=0.45\textwidth]{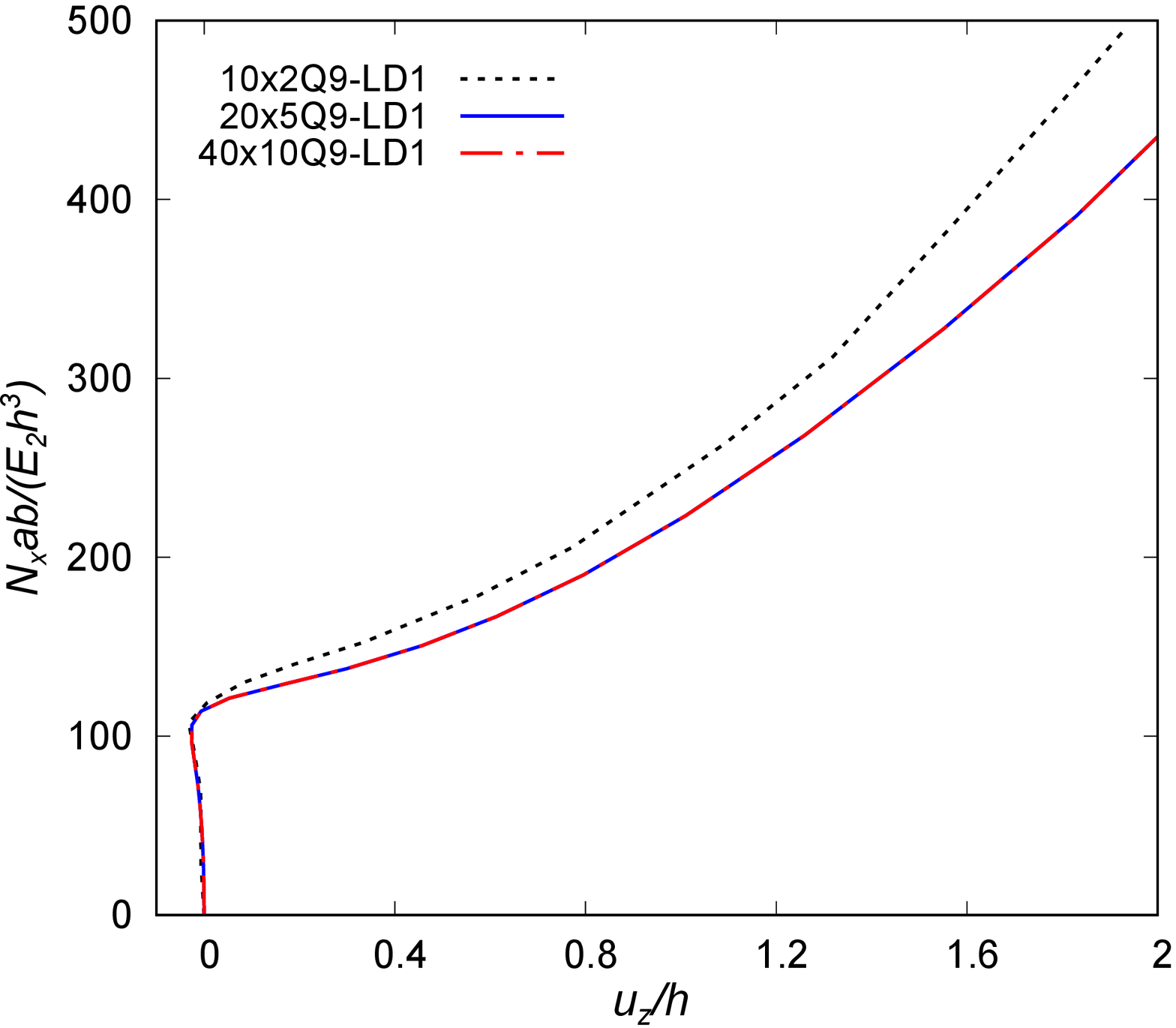}}\hfill
\subfloat[Kinematic expansion\label{ex10-convergez}]{
\includegraphics[width=0.45\textwidth]{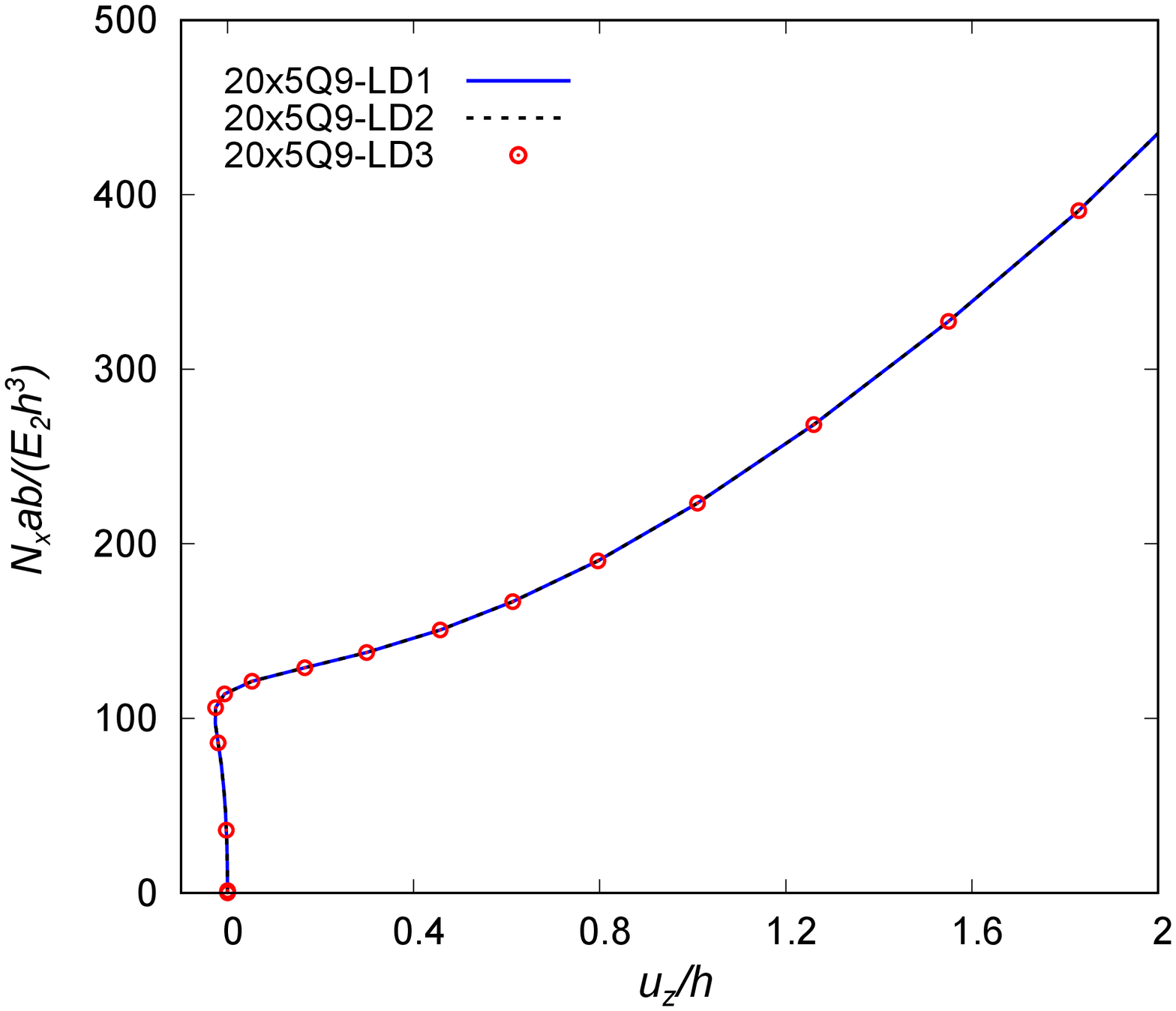}}\hfill
\caption{Convergence analysis for a cross-ply $[0/90]_{2}$ laminate under in-plane compressive line loads in the $x$-axis direction with simply-supported edge conditions.} 
\label{ex10-converge}
\end{figure}

For this composite plate, the convergence analysis of the equilibrium curves is reported in Fig.~\ref{ex10-converge}, which plots the normalized values of the displacement at the center of the plate versus the normalized values of the applied compressive line load. To evaluate the effect of in-plane mesh and kinematic expansion approximations, the finite plate elements are first considered to be 10$\times$2Q9, 20$\times$5Q9, and 40$\times$10Q9 with the fixed LD1 kinematic expansion for each layer. Then, the expansion order along the thickness direction is changed from LD1 to LD3, while the finite plate element is fixed at 20$\times$5Q9. 
Moreover, the transverse displacement values for various CUF plate models and loads are reported in Table \ref{Table004} along with the DOFs. 
Fig.~\ref{ex10-converge}a shows that the convergence is achieved in the lower bound and at least for the 20$\times$5Q9-LD1 plate model. Thus, as evident from Fig. \ref{ex10-converge} and Table \ref{Table004}, the convergence is achieved for the nonlinear response curves when using the 20$\times$5Q9-LD1 plate model. 
\begin{table}[htbp]
\centering
\begin{tabular}{cccccc}
\toprule
\toprule 
\multirow{2}{*}{CUF plate model}  & \multirow{2}{*}{DOFs}  &   \multicolumn{2}{c}{$u_{z}/h$ } \\
\cmidrule(l){3-4} 
 &  & $N_{x}ba/E_{2}h^3=200$ & $N_{x}ba/E_{2}h^3=400$ \\
\midrule
10  $\times$ 2Q9-LD1  &  1575  & 0.731 &  1.618 \\
20  $\times$ 5Q9-LD1  &  6765  & 0.859 &  1.865 \\
40  $\times$ 10Q9-LD1  &  25515  & 0.859 &  1.865 \\
\bottomrule
\bottomrule
\end{tabular} 
\caption{Equilibrium points of nonlinear response curves of a cross-ply $[0/90]_{2}$ laminate under in-plane compressive line loads in the $x$-axis direction with simply-supported edge conditions for different CUF late models and loads- The displacement is calculated at the center of the laminated plate.}\label{Table004}
\end{table}

\begin{figure}[htbp]
\centering
\includegraphics[width=0.68\textwidth,angle=0]{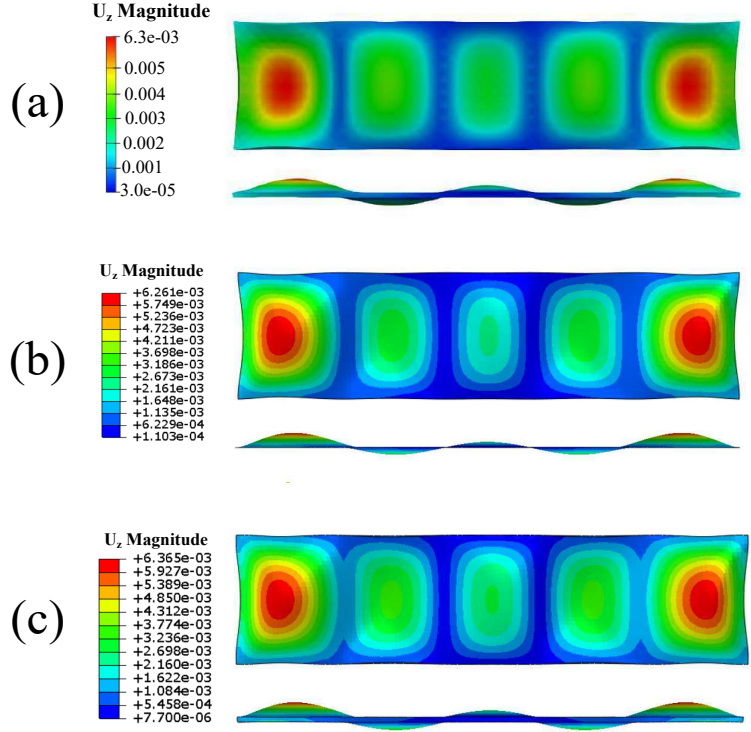}
\hspace{0.00\textwidth}
\caption{Comparison of displacement contours at the fixed load of $\frac{{{N_x}{ba}}}{{E_{2}{h^3}}} = 300$ for a cross-ply $[0/90]_{2}$ laminate under in-plane compressive line loads in the $x$-axis direction with simply-supported edge conditions based on (a) 2D CUF Full NL 20$\times$5Q9+LD1 model, (b) ABQ 2D NL 60$\times$15 S8R model and (c) ABQ 3D NL 60$\times$15$\times$4 C3D20R model.}
\label{ex10acontour}
\end{figure}

Figure \ref{ex10a} shows the equilibrium curves for a cross-ply $[0/90]_2$ composite plate obtained by the 2D CUF Full NL model, ABAQUS (ABQ) 2D shell model and ABQ 3D solid model. As shown in Fig.~\ref{ex10a}, the equilibrium curves obtained by the 2D CUF Full NL model agree well with the ABQ 3D solid model. In contrast, the ABQ 2D shell model predicts accurate results in only the range of small/moderate displacements, while the difference become more remarkable when large displacements are considered. For clarity, in the nonlinear analysis using the ABQ 3D solid model, a fine mesh employing C3D20R elements is used to overcome the mesh instability problem due to the hourglassing. 
Figure \ref{ex10acontour} depicts the deformed configurations and the displacement contours based on the 2D CUF Full NL model (20$\times$5Q9-LD1), ABQ 2D shell model (60$\times$15 S8R) and ABQ 3D solid model(60$\times$15$\times$4 C3D20R) at the fixed load of $\frac{{{N_x}{ba}}}{{E_{2}{h^3}}} = 300$ for the above-mentioned rectangular composite plate. It is clear from this figure that the buckled pattern and the displacement values of different regions predicted by the 2D CUF model have a good consistency with those based on the ABQ models. 

Table \ref{ex10table} shows the displacements values at the fixed load of $\frac{{{N_x}{ba}}}{{E_{2}{h^3}}} = 300$ and the linear buckling loads predicted by the above-mentioned three models for the same rectangular composite plate. 
%
%
%
%
\begin{table}[htbp]
\centering
\begin{tabular}{ccccc}
\toprule
\toprule 
Model &  $u_{z} (mm)$ & Linear Buckling Load $(N/m)$\\
\midrule
2D CUF Full NL 20$\times$5Q9+LD1 & 1.422 & 503360 \\
ABQ 2D NL 60$\times$15 S8R & 1.134 & 497658 \\
ABQ 3D NL 60$\times$15$\times$4 C3D20R &  1.428 & 498976 \\
\bottomrule
\bottomrule
\end{tabular} 
\caption{Comparison of displacements values at the fixed load of $\frac{{{N_x}{ba}}}{{E_{2}{h^3}}} = 300$ and the normalized linear buckling loads for a cross-ply $[0/90]_{2}$ laminate under in-plane compressive line load in the $x$-axis direction with simply-supported edge conditions.}\label{ex10table}
\end{table}
\subsubsection{Cross-ply [0/90]$_{n}$ square laminates with simply-supported edge conditions}
\begin{figure}[htbp]
\centering
\includegraphics[width=0.73\textwidth,angle=0]{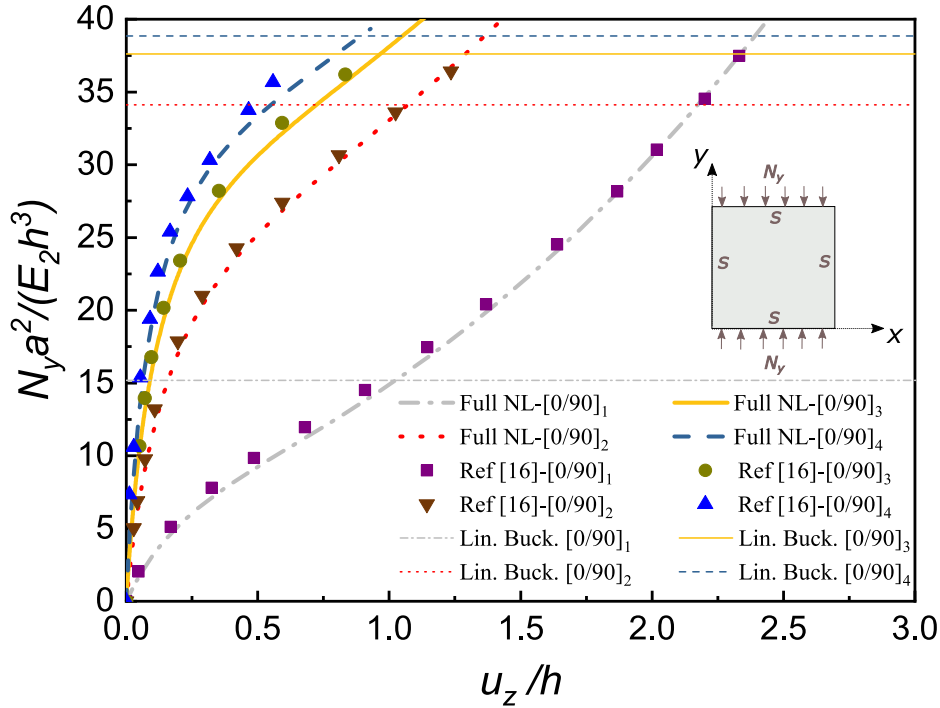}
\hspace{0.00\textwidth}
\caption{Comparison of the equilibrium curves for different cross-ply $[0/90]_{n}$ laminated plates  under in-plane compressive line loads in the $y$-axis direction with simply-supported edge conditions based on 2D CUF Full NL 12$\times$12Q9-LD1 model.}
\label{ex4b}
\end{figure}

The cross-ply $[0/90]_{n}$ composite square plates are investigated in this section \cite{reddy2003mechanics}. The subscript $n$ denotes the number of [0/90] layers. This structure has the width of $a=b=1$~m and the thickness of $h=2$~mm. The plate is subjected to a uniformly distributed in-plane compressive line load in the $y$-axis direction $N_{y}$ (force per unit length, see Fig. \ref{ex4b}) and the edges are simply supported that one set of opposite edges along the $x$-axis direction $x=0,a$ satisfy $v=w=0$, whereas another set of simply-supported opposite edges along the $y$-axis direction $y=0,b$ satisfy $u=w=0$ (see S in Fig. \ref{ex4b}). Table~\ref{ex4properties} provides the material properties of the composite plate. 
%
%
\begin{table}[htbp]
\centering
\begin{tabular}{cccccc}
\toprule
\toprule 
E$_{1}$ (GPa) & $\text{E}_{2}=\text{E}_{3}$ (GPa) & $\text{G}_{12}=\text{G}_{13}$ (GPa) & G$_{23}$ (GPa) & $\nu_{12}=\nu_{13}$ & $\nu_{23}$\\
\midrule
250 &  6.25 & 5.125 &  3.25 &  0.24 & 0.49 \\
\bottomrule
\bottomrule
\end{tabular} 
\caption{Material properties of a $[0/90]_{n}$ composite plate \cite{reddy2003mechanics}.}\label{ex4properties}
\end{table}

\begin{figure}[htbp]
\centering
\includegraphics[width=0.6\textwidth,angle=0]{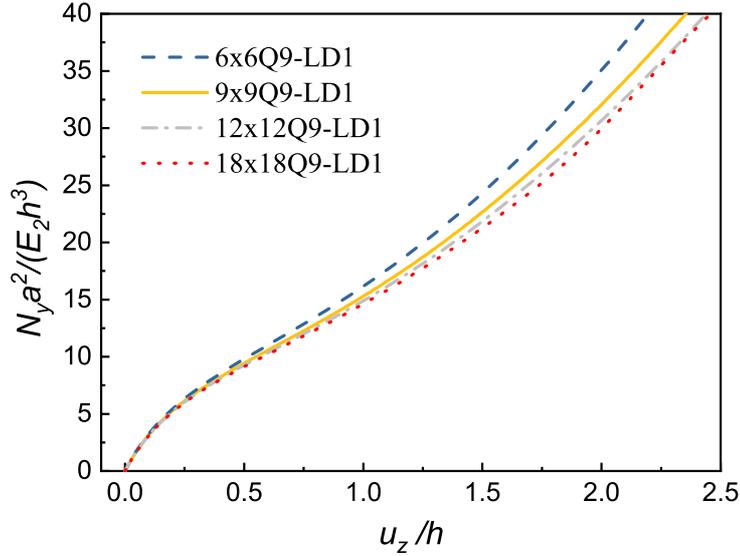}
\hspace{0.00\textwidth}
\caption{Convergence analysis of the in-plane mesh approximation for a cross-ply $[0/90]$ laminate under in-plane compressive line loads in the $y$-axis direction with simply-supported edge conditions.}
\label{ex4b-convergexy}
\end{figure}

For this composite plate structure, the convergence analysis of the equilibrium curves is plotted in Fig.~\ref{ex4b-convergexy}, which plots the normalized values of the displacement in the middle point of the laminate versus the normalized values of the applied compressive load. 
Moreover, the transverse displacement values for various CUF plate models and loads are reported in Table \ref{Table005} along with the DOFs. 
\begin{table}[htbp]
\centering
\begin{tabular}{cccccc}
\toprule
\toprule 
\multirow{2}{*}{CUF plate model}  & \multirow{2}{*}{DOFs}  &   \multicolumn{2}{c}{$u_{z}/h$ } \\
\cmidrule(l){3-4} 
 &  & $N_{y}a^2/E_{2}h^3=15$ & $N_{y}a^2/E_{2}h^3=35$ \\
\midrule
6   $\times$ 6Q9-LD1  &  1521  & 0.90 & 1.97  \\
9   $\times$ 9Q9-LD1  &  3249  & 0.97 &  2.11 \\
12  $\times$ 12Q9-LD1  &  5625  & 1.01 & 2.19  \\
18  $\times$ 18Q9-LD1  &  12321 & 1.03 & 2.20  \\
\bottomrule
\bottomrule
\end{tabular} 
\caption{Equilibrium points of nonlinear response curves of a cross-ply $[0/90]$ laminate under in-plane compressive line loads in the $y$-axis direction with simply-supported edge conditions for different CUF plate models and loads. The displacement is calculated at the center of the laminated plate.}\label{Table005}
\end{table}
As shown in Fig. \ref{ex4b-convergexy} and Table \ref{Table005}, the convergence is achieved at least for the 12$\times$12Q9-LD1 model. 

The equilibrium curves obtained by the 2D CUF Full NL model and solutions in the available literature are compared in Fig.~\ref{ex4b}. In this figure, the horizontal lines show the corresponding linear buckling load by the CUF method. It is clear in Fig.~\ref{ex4b} that the equilibrium curves predicted by the 2D CUF Full NL model agree well with those available in the literature \cite{reddy2003mechanics}. The results show the fact that by assuming a constant value for the plate thickness, increasing the layer number of the composite plate results in higher structural stiffness and load-carrying capacity of the plate. Moreover, the linear buckling strength is increased significantly for the $[0/90]_{4}$ plate with 8-layers compared to the $[0/90]_{1}$ plate with 2-layers. Also, it can be understood that the linear buckling strength of the $[0/90]$ composite plate is dramatically lower than all other investigated composite plates.
It is noted from Fig. \ref{ex4b} that no exact buckling load exists for the $[0/90]_n$ composite plate structure based on the CUF Full NL plate model. This is because the antisymmetric composite laminate is under in-plane compressive loads. The buckling load predicted by the linear buckling analysis is much higher than that based on the Full NL plate model. Therefore, the linear buckling analysis cannot be utilized to calculate the buckling load of the antisymmetric plate structure due to the curvature introduced by the in-plane compressive loads.
\subsubsection{Angle-ply [45/-45]$_{s}$ laminate with simply-supported edge conditions}
\begin{figure}[htbp]
\centering
\includegraphics[width=0.4\textwidth,angle=0]{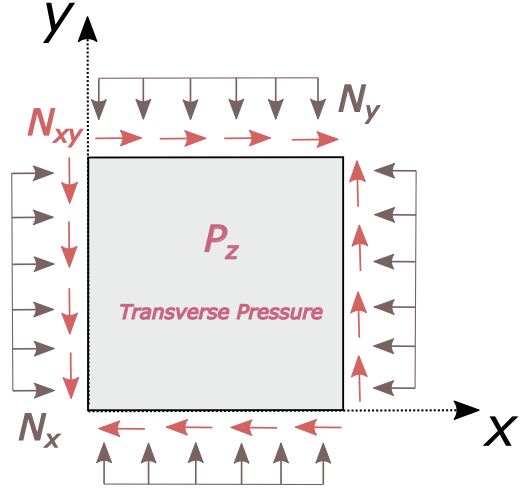}
\hspace{0.00\textwidth}
\caption{Combined loading of a laminated composite plate: positive in-plane shear, in-plane compression, and the uniform transverse pressure.}
\label{ex7schematic}
\end{figure}

This analysis case deals with angle-ply $[45/-45]_{s}$ composite square plate under the combined loading \cite{han2006postbuckling}. The width of the plate is $a=b=0.25$~m and the thickness is $h=2.5$~mm. Different loading cases are assumed for the composite plate such as the combination of uniformly distributed in-plane compressive bi-axial line loads of $N_{x}$ and $N_{y}$ ($N_{x}=N_{y}$ in the current example), the in-plane shear load of $N_{xy}=N_{x}$, and the uniform transverse pressure of $P_{z}=0.1N_{x}$. The edge conditions are simply-supported in such a way that only the transverse deflections are restrained at the edges. The material properties and a schematic view of the loading conditions are shown in Table~\ref{ex7properties} and Fig.~\ref{ex7schematic}, respectively. 
%
%
\begin{table}[htbp]
\centering
\begin{tabular}{cccccc}
\toprule
\toprule 
E$_{1}$ (GPa) & $\text{E}_{2}=\text{E}_{3}$ (GPa) & $\text{G}_{12}=\text{G}_{13}$ (GPa) & $\nu_{12}=\nu_{13}$ \\
\midrule
206.9 &  5.2 & 2.6  &  0.25 \\
\bottomrule
\bottomrule
\end{tabular} 
\caption{Material properties of a 4-layer $[45/-45]_{s}$ composite plate \cite{han2006postbuckling}.}\label{ex7properties}
\end{table}

\begin{figure}[htbp]
\centering
\subfloat[Positive shear\label{ex7lambda-convergexy-pos}]{
\includegraphics[width=0.44\textwidth]{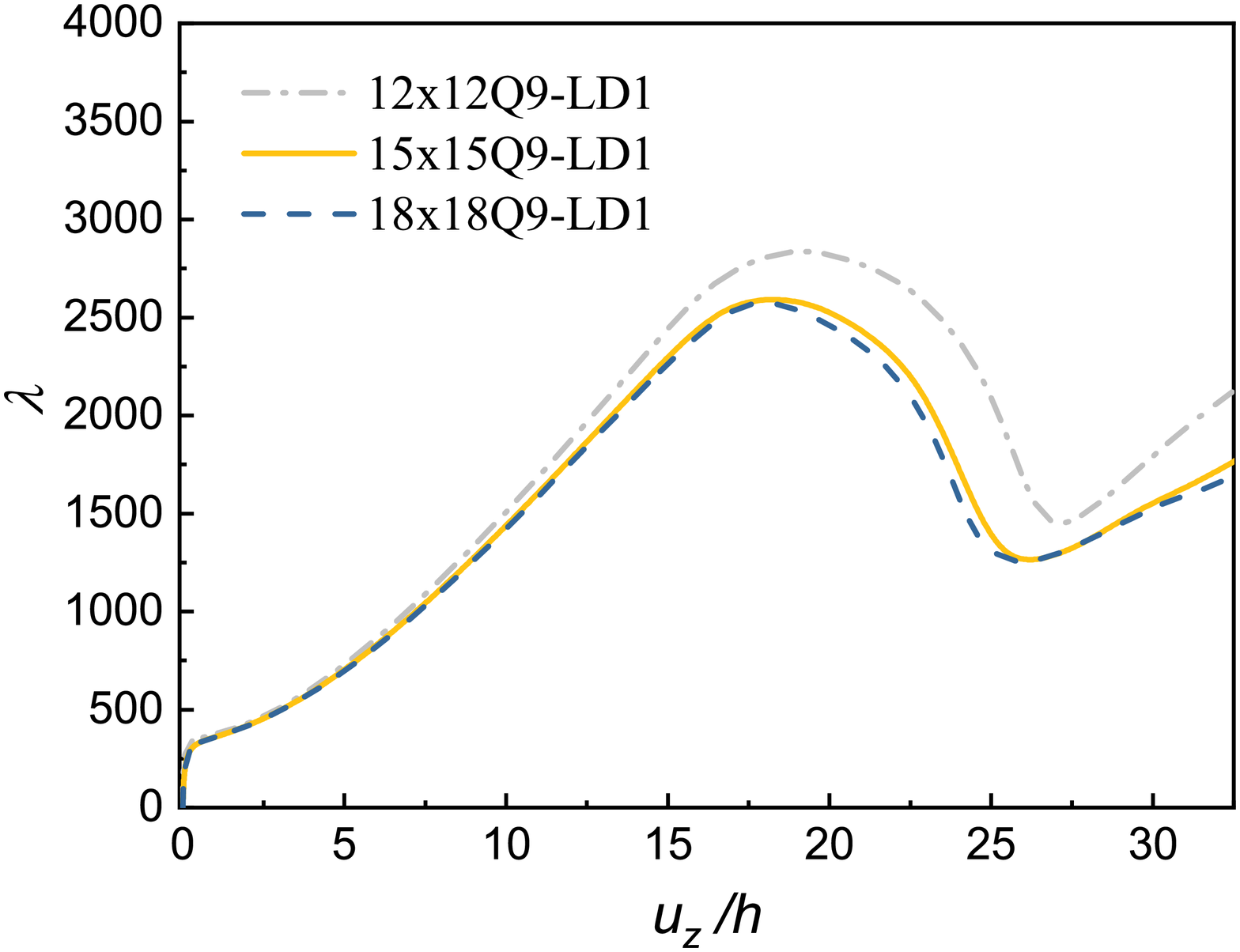}}\hfill
\subfloat[Negative shear\label{ex7lambda-convergexy-neg}]{
\includegraphics[width=0.44\textwidth]{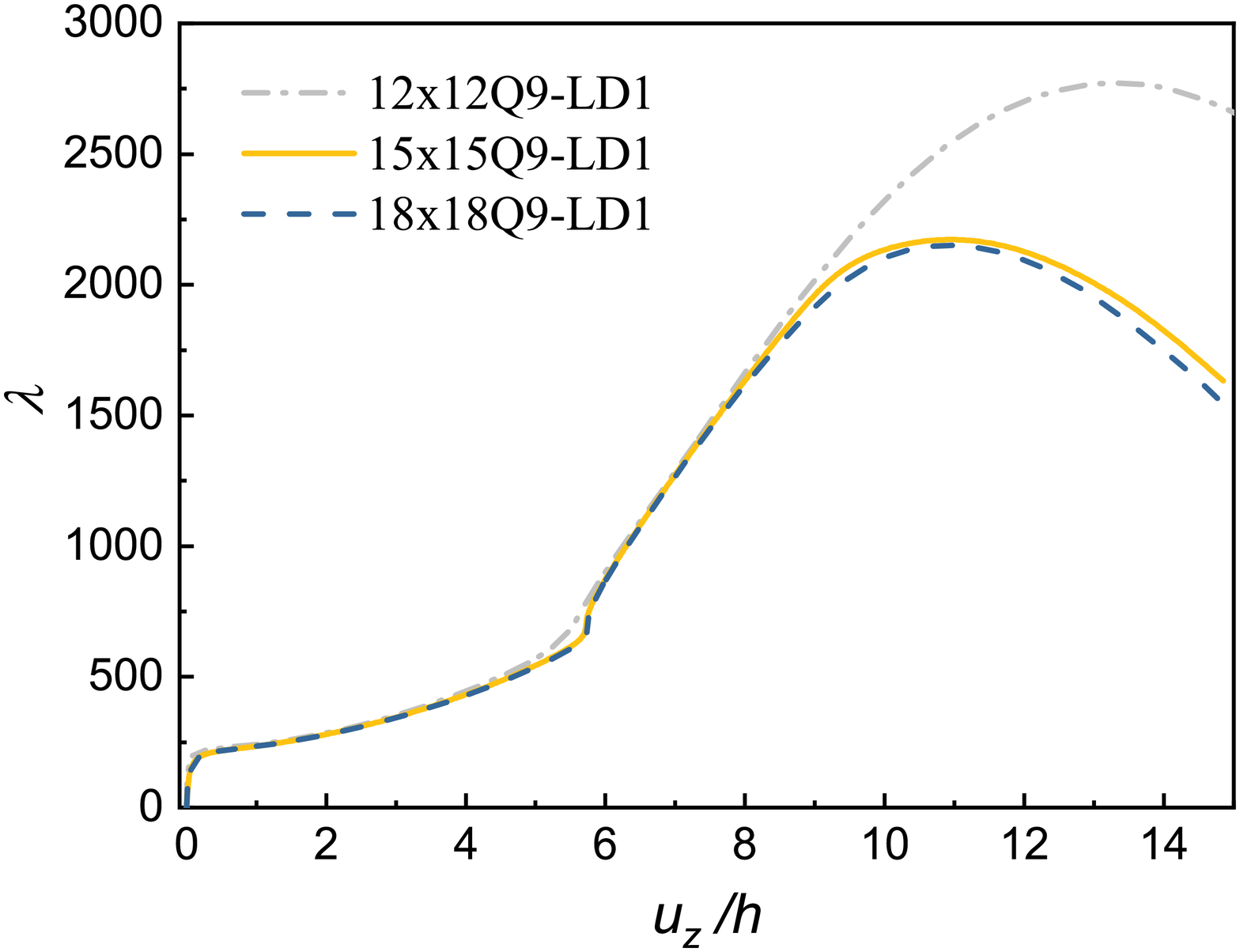}}\hfill
\caption{Convergence analysis for an angle-ply $[45/-45]_{s}$ laminate under the combined loading with simply-supported edge conditions.} 
\label{ex7lambda-convergexy}
\end{figure}

For this composite structure, the convergence analysis of the equilibrium curves is provided in Fig.~\ref{ex7lambda-convergexy}, which plots the normalized values of the displacement in the middle point of the plate versus the values of the loading factor ($\lambda$). 
It is evident from Fig. \ref{ex7lambda-convergexy} that the convergence is reached at least for the 15$\times$15Q9-LD1 plate model.

\begin{figure}[htbp]
\centering
\includegraphics[width=0.7\textwidth,angle=0]{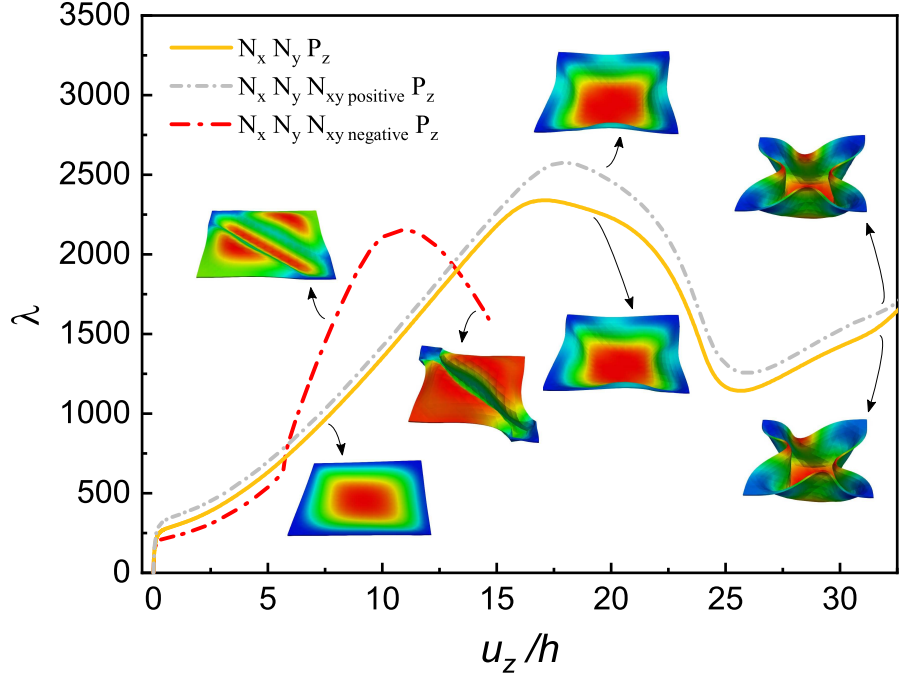}
\hspace{0.00\textwidth}
\caption{Comparison of the equilibrium curves based on the values of loading factor $\lambda$ for an angle-ply $[45/-45]_{s}$ laminate under different combined loadings with simply-supported edge conditions.}
\label{ex7lambda}
\end{figure}

Fig.~\ref{ex7lambda} illustrates how the deflection at the middle of the laminate varies by increasing the loading after the bifurcation point. The results show the fact that the direction of the applied shear loading plays a pivotal role in the post-buckling behaviour of the angle-ply composite plate. Specially, in contrast with the negative shear loading, the angle-ply plate with positive shear loading shows higher rigidity and load-carrying capacity, which will be further displayed below.

\begin{figure}[htbp]
\centering
\includegraphics[width=0.7\textwidth,angle=0]{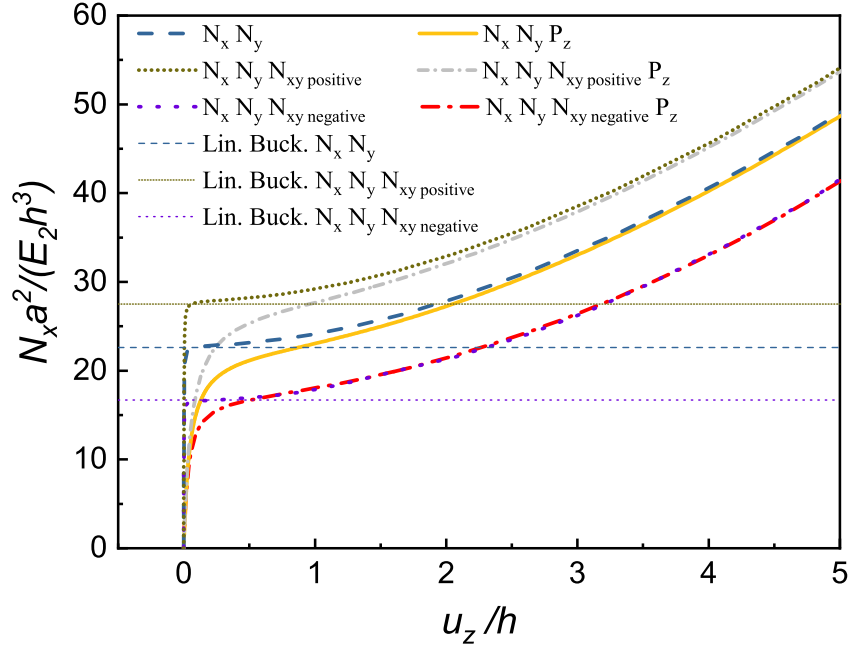}
\hspace{0.00\textwidth}
\caption{Comparison of the equilibrium curves for an angle-ply $[45/-45]_{s}$ laminate under different combined loadings with simply-supported edge conditions.}
\label{ex7}
\end{figure}

The equilibrium curves for this angle-ply composite plate subjected to different combined loadings are shown in Fig.~\ref{ex7}, which plots the normalized values of the displacement in the middle point of the plate versus the normalized values of the applied compressive line load in the $x$-axis direction. The horizontal lines in this figure display the corresponding linear buckling loads predicted by the CUF method. 
It is found from Fig. \ref{ex7} that for this symmetric composite structure, the buckling loads predicted by the linear buckling analysis are almost the same as those based on CUF Full NL plate model, even if there exists uniform transverse pressure applied to the plate. Thus, the linear buckling analysis can be exploited to first predict the buckling load of the symmetric composite structure. 
In addition, the linear buckling load of the angle-ply plate with positive shear is higher than other loading cases, which demonstrates the previously mentioned fact that the angle-ply plate with positive shear loading has higher rigidity and load-carrying capacity.
Finally, it is noted that as the transverse pressure is relatively small compared with the in-plane loads, the equilibrium curves with transverse pressure gradually approach those without transverse pressure when continuously increasing the loading.
\subsubsection{Cross-ply [0/90] square laminate with different edge conditions}
The effect of different plate edge conditions on the post-buckling nonlinear response of the cross-ply composite plate under in-plane compressive load is presented for the final post-buckling example. This problem deals with a cross-ply $[0/90]$ square plate with different edge conditions \cite{liew2006postbuckling}. This 2D model has the width of $a=b=1$~m and the thickness of $h=1$~cm. In the first analysis case, the plate is subjected to a uniformly distributed in-plane compressive line load in the $x$-axis direction $N_{x}$, while in the second analysis case, the plate is subjected to a uniformly distributed in-plane compressive line load in the $y$-axis direction $N_{y}$. The edge conditions are SSSS, SSCC, SSSC, SSFC, and SSFS. The letters ``S", ``C", and ``F" indicate simply-supported, clamped, and free edge conditions. It should be noted that the third and fourth letters of the boundary condition refer to $y=b$ and $y=0$. Furthermore, the clamped edge conditions satisfy $u=v=w=0$ at the corresponding edge and the simply-supported edge conditions satisfy $v=w=0$ at $x=0,a$, or $u=w=0$ at $y=0,b$. Material properties for the bove-mentioned composite plate are shown in Table~\ref{ex6properties}.   
%
%
\begin{table}[htbp]
\centering
\begin{tabular}{cccccc}
\toprule
\toprule 
E$_{1}$ (GPa) & $\text{E}_{2}=\text{E}_{3}$ (GPa) & $\text{G}_{12}=\text{G}_{13}$ (GPa) &   G$_{23}$ (GPa) & $\nu_{12}=\nu_{13}$ \\
\midrule
220 &  5.5 & 3.3  &  2.75  & 0.25 \\
\bottomrule
\bottomrule
\end{tabular} 
\caption{Material properties of a 2-layer $[0/90]$ composite plate \cite{liew2006postbuckling}.}\label{ex6properties}
\end{table}

\begin{figure}[htbp]
\centering
\includegraphics[width=0.6\textwidth,angle=0]{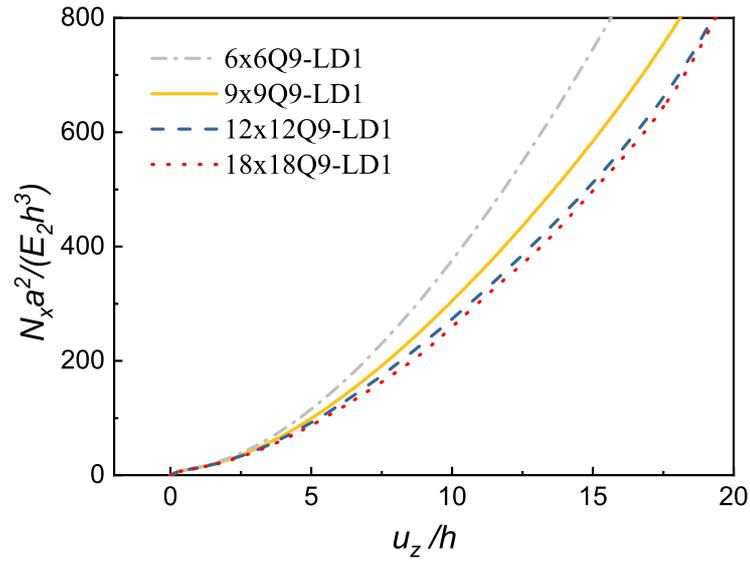}
\hspace{0.00\textwidth}
\caption{Convergence analysis of the in-plane mesh approximation for a cross-ply $[0/90]$ laminate under in-plane compressive loads in the $x$ axis direction with simply-supported edge conditions.}
\label{ex6a-convergexy}
\end{figure}

\begin{figure}[htbp]
\centering
\includegraphics[width=0.8\textwidth,angle=0]{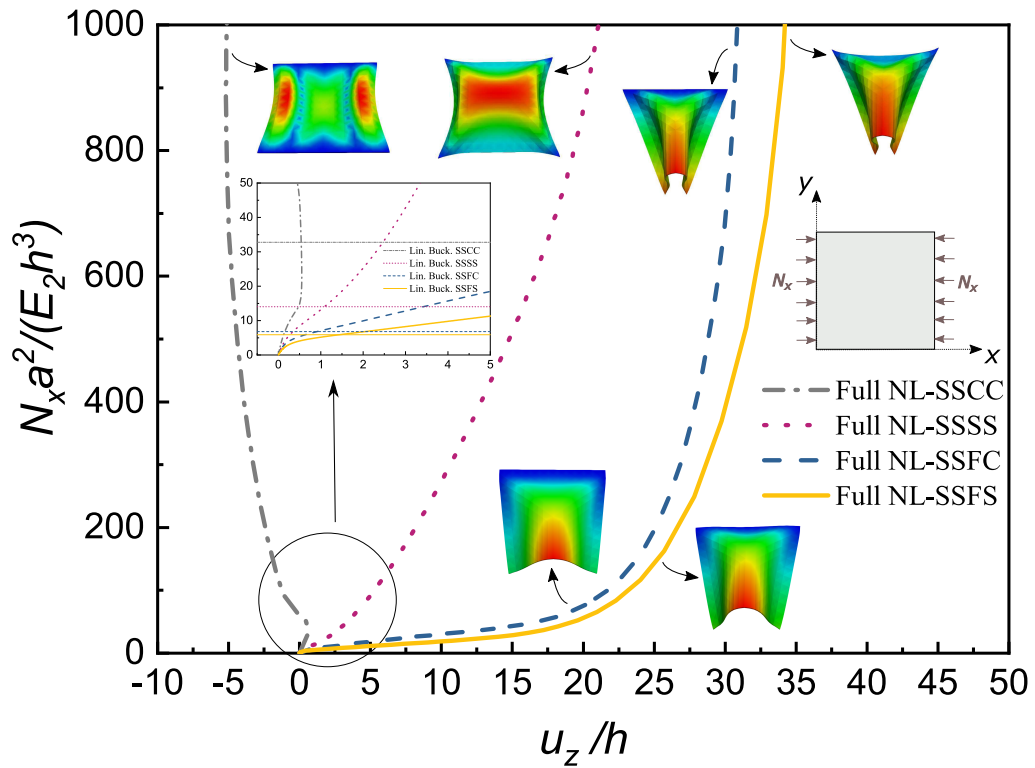}
\hspace{0.00\textwidth}
\caption{Comparison of the equilibrium curves for a cross-ply $[0/90]$ laminate under in-plane compressive line loads in the $x$-axis direction with different edge conditions based on 2D CUF Full NL 12$\times$12Q9-LD1 model.}
\label{ex6a}
\end{figure}

\begin{figure}[htbp]
\centering
\includegraphics[width=0.8\textwidth,angle=0]{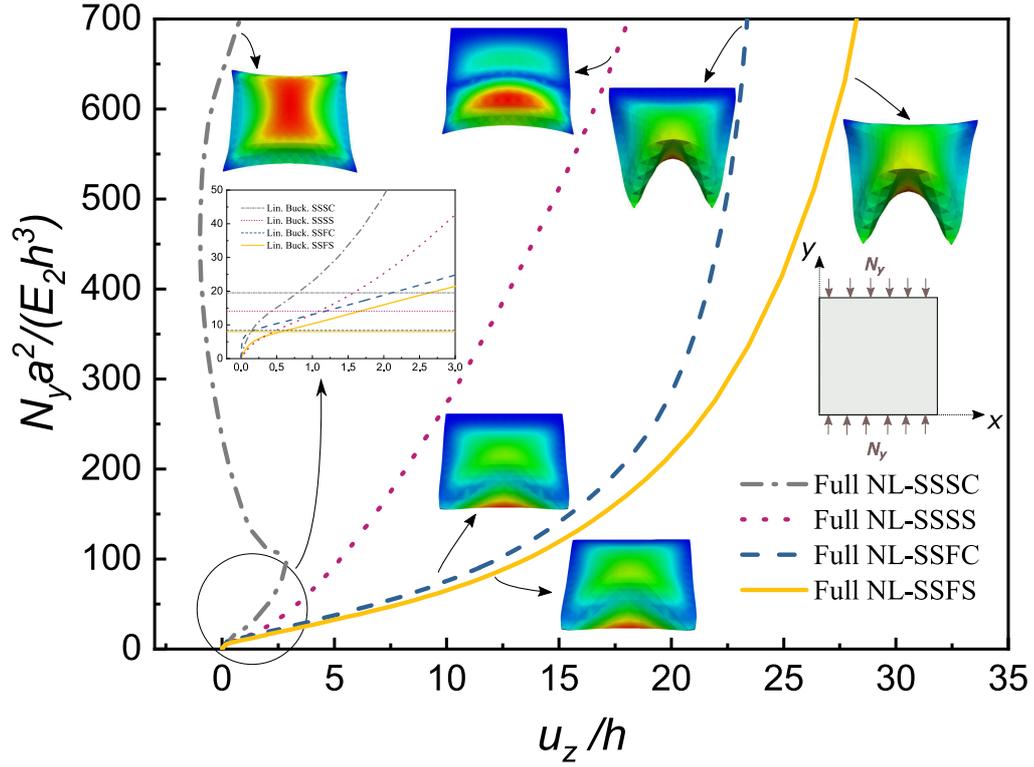}
\hspace{0.00\textwidth}
\caption{Comparison of the equilibrium curves for a cross-ply $[0/90]$  laminate under in-plane compressive line loads in the $y$-axis direction with different edge conditions based on 2D CUF Full NL 12$\times$12Q9-LD1 model.}
\label{ex6b}
\end{figure}

For this cross-ply composite plate with simply-supported edge conditions, the convergence analysis of the equilibrium curves for the in-plane compressive loads in the $x$-axis direction is provided in Fig.~\ref{ex6a-convergexy}.
This figure shows that the convergence is obtained at least for the 12$\times$12Q9-LD1 plate model. The equilibrium curves based on this convergence plate model are investigated for SSSS, SSCC, SSSC, SSFC, and SSFS edge conditions in Figs.~\ref{ex6a} and \ref{ex6b} to evaluate the effect of different edge conditions on the nonlinear response of the cross-ply composite plate. The horizontal lines in these two figures' enlarged views indicate the corresponding linear buckling loads by the CUF method. The results show that the load-carrying capacity of the composite plate with the clamped edge conditions is higher than other investigated edge conditions. Furthermore, it can be understood that the presence of a free edge reduces the buckling strength significantly. The current method can also predict the nonlinear response of the composite plate beyond the limit load and the snap-through instability. 
%
%
%
\newpage
\section{Conclusions}\label{Z}
In this work, large deflection and post-buckling analyses of laminated composite plates have been investigated by employing the Carrera Unified Formulation (CUF) and the layerwise (LW) approach based on Lagrange expansion. The path-following Newton-Raphson linearization method has been adopted to compute the full geometrically nonlinear plate problems. Different two-dimensional (2D) composite plate structures subjected to large deflections/rotations and post-buckling have been analyzed and the numerical results have been compared with solutions in the available literature. Furthermore, the linear buckling load of the composite plates has been calculated for the post-buckling cases. The effects of different parameters, such as the stacking sequence, number of layers, loading way, and edge conditions, have been investigated and discussed in detail. 
The results have demonstrated that:
\begin{itemize}
\item The equilibrium curves obtained by the CUF linear and Full NL models agree well with those in the available literature or the ABAQUS 3D solid model solutions;       
\item Increasing the layer number of the composite plates results in the higher buckling strength and load-carrying capacity of the composite structures; 
\item For the angle-ply laminate subjected to the combined loading (in-plane shear and bi-axial compression), the direction of the applied shear plays a pivotal role in the post-buckling behaviour of the composite plate, and the angle-ply plate with positive shear loading shows higher rigidity and load-carrying capacity;
\item The buckling strength and the load-carrying capacity of the composite plates with the clamped edge conditions are higher compared to other investigated edge conditions, and the presence of a free edge reduces the buckling strength significantly;
\item The linear buckling analysis cannot be utilized to calculate the buckling load of the antisymmetric structure. However, the buckling loads predicted by the linear buckling analysis for symmetric composite structures are almost the same as those based on CUF Full NL plate model.
\end{itemize}	

Further developments of the proposed methodology are being studied, such as a non-linear local analysis and a localized buckling with the advantage of coupling the global/local approach with optimization tools to reduce computation time. Furthermore, the same non-linear methodology will also be adopted to perform dynamic analyses. Another important topic under development is the use of CUF in civil problems, such as the analysis of reinforced concrete structures. The adoption of an LW approach for studying the nonlinear behaviour of the section once the concrete cracks have some potential and advantages on the accuracy of the results and processing times.
\section*{Declaration of competing interest}
Authors declare that they have no known competing financial interests or personal relationships that could appear to influence the work reported in this paper.
\section*{Author contributions}
\textbf{E. Carrera}: Conceptualization, Funding acquisition, Methodology.
\textbf{R. Azzara}: Investigation, Visualization, Writing - original draft.
\textbf{E. Daneshkhah}: Investigation, Visualization, Writing - original draft.
\textbf{A. Pagani}: Supervision, Software, Writing - review $\&$ editing.
\textbf{B. Wu}: Visualization, Writing - review $\&$ editing.
\section*{Acknowledgments}
This work was supported by the European Union’s Horizon 2020 Research and Innovation Programme under the Marie Skłodowska-Curie Actions, Grant No. 896229 (B.W.).
\bibliographystyle{unsrt}
\bibliography{NL_plate}
\end{document}